\documentclass[reqno, centertags, 11pt]{amsart}
\usepackage{amsmath,amsthm,amsfonts,amssymb,graphicx}

\newcommand{\nl}{\par\noindent}
\newcommand{\Lu}{\L uka\-sie\-wicz}

\newcommand{\Tr}{{\tt tr}}
\def\part #1{\left(#1\right)}
\newcommand{\parq}[1]{\left[#1\right]}
\newcommand{\parg}[1]{\left\{#1\right\}}
\newcommand{\um}{\dfrac{1}{2}}
\newcommand{\FM}{Form^{\mathcal{L}}}

\newcommand{\Prob}{{\tt p}}
\newcommand{\QNot}{{\tt Not}}
\newcommand{\QSNot}{\sqrt{\tt Not}}
\newcommand{\QAnd}{{\tt And}}
\newcommand{\QOr}{{\tt Or}}
\newcommand{\NOT}{{\tt NOT}}
\newcommand{\SNOT}{\sqrt{\tt NOT}}
\newcommand{\AND}{{\tt AND}}
\newcommand{\IAND}{{\tt IAND}}
\newcommand{\OR}{{\tt OR}}
\newcommand{\Qum}{{\tt Qum}}
\newcommand{\Qub}{{\tt Qub}}
\newcommand{\ket}[1]{\vert {#1} \rangle}
\newcommand{\dket}[1]{\left\Vert {#1} \right\rangle \! \rangle}

\newcommand{\N}{\mathbb{N}}
\newcommand{\R}{\mathbb{R}}
\newcommand{\C}{\mathbb{C}}

\newcommand{\qclr}{\models_{_{^{\sqrt{\neg}}\rm{QCL}}}}

\newtheorem{theorem}{Theorem}[section]
\newtheorem{lemma}{Lemma}[section]

\newtheorem{corollary}{Corollary}[section]
\newtheorem{definition}{Definition}[section]
\newtheorem{example}{Example}[section]

\numberwithin{equation}{section}

\begin{document}

\title[Quantum Computational Logics]{\textbf{Quantum Computational Logics. A Survey.}}
\author[M.~L.~Dalla Chiara]{Maria Luisa Dalla Chiara}
\address[M.~L.~Dalla Chiara]{Dipartimento di Filosofia,
    Universit\`a di Firenze,
    via Bolognese 52, I-50139 Firenze, Italy}
\email{dallachiara@unifi.it}
\author[R.~Giuntini]{Roberto Giuntini}
\address[R.~Giuntini]{Dipartimento di Scienze Pedagogiche e Filosofiche,
    Universit\`a di Cagliari,
    via Is Mirrionis 1, I-09123 Cagliari, Italy}
\email{giuntini@unica.it}
\author[R. Leporini]{Roberto Leporini}
\address[R. Leporini]{Dipartimento di Informatica, Sistemistica e Comunicazione (DISCo),
    Universit\`a degli Studi di Milano--Bicocca,
    via Bicocca degli Arcimboldi 8, I-20126 Milano, Italy}
\email{leporini@disco.unimib.it}

\date{}
\maketitle

\begin{abstract}
Quantum computation has suggested new forms of quantum logic,
called \emph{quantum computational logics} (\cite{CDCGL-02}). The
basic semantic idea is the following: the meaning of a sentence is
identified with a \emph{quregister}, a system of \emph{qubits},
representing a possible \emph{pure state} of a compound quantum
system. The generalization to \emph{mixed states}, which might be
useful to analyse \emph{entanglement-phenomena}, is due to Gudder
(\cite{Gu-02}). Quantum computational logics represent non standard
examples of \emph{unsharp quantum logic}, where the
non-contradiction principle is violated, while conjunctions and
disjunctions are strongly non-idempotent. In this framework, any
sentence $\alpha$ of the language gives rise to a \emph{quantum
tree}: a kind of \emph{quantum circuit} that transforms the
quregister associated to the atomic subformulas of $\alpha$ into
the quregister associated to $\alpha$.
\end{abstract}

\section{Introduction}

Quantum computation has suggested new forms of quantum logic that
have been called \emph{quantum computational logics}. The main
difference between orthodox quantum logic (first proposed by
Birkhoff and von Neumann \cite{BVN-36}) and quantum computational
logics concerns a basic semantic question: how to represent the
\emph{meanings} of the sentences of a given language? The answer
given by Birkhoff and von Neumann is the following: the meanings
of the elementary experimental sentences of quantum theory (QT)
have to be regarded as determined by convenient sets of
\emph{states} of quantum objects. Since these sets should satisfy
some special closure conditions, it turns out that, in the
framework of orthodox quantum logic, sentences can be adequately
interpreted as \emph{closed subspaces} of the Hilbert space
associated to the physical systems under investigation. The answer
given in the framework of quantum computational logics is quite
different: meanings of sentences are represented by
\emph{information quantities}, a kind of abstract objects that are
described in the framework of \emph{quantum information theory}.

\section{From classical to quantum information}

As is well known, the unit of measurement in classical information
theory is the \emph{bit}: one \emph{bit} measures the
information quantity that can be either transmitted or received
whenever one chooses one element from a set consisting of two
distinct elements (say, from the set $\parg{0,1}$). From an
intuitive point of view, both the objects $0$ and $1$ can be
imagined as a well determined \emph{state} of a classical
physical system (for instance, the state of a tape cell in a given
machine).

Let us now refer to a quantum computational context, where
information is supposed to be elaborated and transmitted by means
of a quantum system. According to the standard axiomatization of
QT, the \emph{pure states} of our system are mathematically
represented by unit vectors in a convenient Hilbert space
$\mathcal H$. Let us refer to the simplest situation, where our
Hilbert space $\mathcal H$ has dimension 2; hence $\mathcal H=
\C^2$. In such a case $\mathcal H$ will have a \emph{basis}
consisting of two unit elements, and any vector of the space will
be representable as a \emph{superposition} of the two
basis-elements. In quantum computation, it is customary to use
Dirac's notation. Accordingly, the vectors of $\mathcal H$ are
indicated by $\ket{\psi}$, $\ket{\varphi}$,... ; while the
basis-elements are denoted by $\ket{0}$, $\ket{1}$. As a
consequence, for any unit vector $\ket{\psi}$ we will have:
$$ \ket{\psi}= a_0\ket{0} + a_1\ket{1}, $$
where the coefficients $a_0,a_1$ are complex numbers
(also called amplitudes) such that:
$$ |a_0|^2+|a_1|^2=1. $$

Let us now try and apply such a formalism to a quantum information
theory. The basic idea is to generalize the concept of
\emph{bit}, by introducing the notion of \emph{qubit} or
\emph{quantum bit}. A \emph{qubit} is a unit vector in the
Hilbert space $\C^2$. Thus any \emph{qubit} will have the form
$\ket{\psi}= a_0\ket{0} + a_1\ket{1}.$ The interpretation is
determined by an axiom of QT that is usually called the \emph{Born
rule}. Suppose that (like in the classical case) the pure states
$\ket{0}$ and $\ket{1}$ represent two \emph{maximal} (and
\emph{precise}) pieces of information. Then the
superposition-state $\ket{\psi}= a_0\ket{0} + a_1\ket{1}$ will
represent an information that involves a certain degree of
\emph{uncertainty}. In particular, the number $|a_0|^2$ will
correspond to the probability-value of the information described
by the basic state $\ket{0}$; while $|a_1|^2$ will correspond to
the probability-value of the information described by the basic
state $\ket{1}$.

In this context, it makes sense to imagine $\ket{\psi}$ as an
\emph{epistemic state} that stocks two precise pieces of
information \emph{in parallel}: the information $\ket{0}$ and the
information $\ket{1}$.

Let us now consider a situation characterized by many \emph{bits}
or \emph{qubits}. As is well known, in classical information
theory, a system consisting of $n$ \emph{bits} is naturally
represented by a sequence of $n$ elements belonging to the set
$\parg{0,1}$ (i.e. as an element of the set $\parg{0,1}^n$). In
the framework of quantum computation, it is convenient to adopt
the tensor product formalism, which is used in quantum theory in
order to represent compound physical systems. Suppose a
two-particle quantum system:
$$ \frak S = \frak {S}_1 + \frak {S}_2. $$
For instance, $\frak {S}_1$ and $\frak {S}_2$ might
correspond respectively to the two electrons in a given helium
atom. In such a case, the Hilbert space $\mathcal H^{\frak S}$
associated to $\frak S$ will be the tensor product $\mathcal
H^{\frak {S}_1}\otimes \mathcal H^{\frak {S}_2}$ of the two
Hilbert spaces $\mathcal H^{\frak {S}_1}$ and $\mathcal H^{\frak
{S}_2}$, that are associated to $\frak {S}_1$ and $\frak {S}_2$,
respectively. Thus any pure state of $\frak S$ will be a unit
vector in the space $\mathcal H^{\frak S}$.

A particularly interesting case is represented by those vectors
$\ket{\psi}$ of $\mathcal H^{\frak S}$ that can be expressed as
the tensor product of two vectors $\ket{\psi_1}$ and $\ket{\psi_2}$,
belonging to $\mathcal H^{\frak {S}_1}$ and to $\mathcal H^{\frak
{S}_2}$, respectively. In other words:
$$ \ket{\psi}= \ket{\psi_1} \otimes \ket{\psi_2}. $$
In such cases, one usually speak of
\emph{factorized states}. It is worthwhile recalling that not
all vectors of $\mathcal H^{\frak S}$ can be expressed in such a
simple form.

How to represent, in this framework, a system consisting of $n$
\emph{qubits}? It seems quite natural to describe our system as
the pure state of a compound physical system consisting of $n$
quantum objects. On this basis our $n$-\emph{qubit} system can
be identified with a unit vector of the product space
$$ \underbrace{\C^2 \otimes\ldots\otimes\C^2 }_{n-times}. $$

Instead of $\underbrace{\C^2 \otimes\ldots\otimes\C^2 }_{n-times}$
we will also write: $\otimes^n \C^2$. Particularly interesting
examples will be represented by the vectors of $\otimes^n\C^2 $
whose form is:
$$ \ket{x_1}\otimes \ldots \otimes \ket{x_n}, $$
where each $\ket{x_i}$ is an element of the basis of $\C^2 $
(i.e., $\ket{x_i} = \ket{0}$ or $\ket{x_i} = \ket{1}$). One can
prove that the set of all vectors having this form represents a
basis for the product space $\otimes^n\C^2$.

How can we deal with the concept of \emph{quantum computation},
in this framework? The basic idea is to describe a computation by
means of that kind of process that corresponds to the {dynamic
evolution} of a quantum system. Suppose a physical system $\frak
S$, whose pure state at time $t_0$ is the vector $\ket{\psi(t_0)}$
(where $\ket{\psi(t_0)}$ belongs to the Hilbert space $\mathcal
H^{\frak S}$ associated to $\frak S$). Owing to Schr\"odinger's
equation, for any time $t$ (where either $t\leq t_0$ or $t_0 \leq
t$), there exists an operator $U_{[t_0,t]}$ that determines the
state of the system at time $t$ as a function of the state of the
system at time $t_0$. In other words:
$$ \ket{\psi(t)}=U_{[t_0,t]}\ket{\psi(t_0}. $$
Any operator $U_{[t_0,t]}$ of this kind is \emph{unitary}.
Hence, our operator preserves the length of the vectors and the
orthogonality relation. Further it is \emph{reversible}: one can
go from $\ket{\psi(t_0)}$ to $\ket{\psi(t)}$ and viceversa, without
any dissipation of information.

On this ground, it makes sense to represent a quantum computation
by means of convenient unitary operators assuming arguments and
values in particular sets of \emph{qubit} systems. Since
\emph{qubits} are generally superposition-states, one obtains
some typical \emph{parallel configurations}.

\section{Qubits, Quregisters and Qumixs}

As we have seen, qubits and qubit-systems (also called
\emph{quregisters}) correspond to pure states, which are
\emph{maximal} pieces of information of the observer about the
quantum system under investigation. In other words, one is dealing
with a kind of information that cannot be consistently extended to
a richer knowledge (expressed in the same language). In many
concrete situations it may be interesting to consider also
\emph{mixed states} (or \emph{mixtures}), describing pieces of
information that are not generally maximal. According to the
standard axiomatization of QT such states are mathematically
represented by \emph{density operators} $\rho$ of the Hilbert
space $\mathcal H$ (associated to the system). Any pure state
$\ket{\psi}$ corresponds to a limit-case of a density operator:
the projection $P_{\ket{\psi}}$ onto the one-dimensional closed
subspace determined by the vector $\ket{\psi}$. Representing
quantum information by density operators turns out to be important
in order to deal with \emph{entanglement-phenomena}, which play a
fundamental role in teleportation and in quantum cryptography.

We will now sum up some basic formal definitions of quantum
computation. Consider the two--dimensional Hilbert space $\C^2$
(where any vector $\ket{\psi}$ is represented by a pair of complex
numbers). Let $\mathcal{B}^{(1)}=\{ \ket{0}, \ket{1} \}$ be the
canonical orthonormal basis for $\C^2$, where $\ket{0} = (1,0)$
and $\ket{1} = (0,1)$.

\begin{definition}(Qubit). \label{de:qubit} \nl
A \emph{qubit} is a unit vector $\ket{\psi}$ of the Hilbert space
$\C^2$.
\end{definition}
As we have seen, from an intuitive point of view, any qubit
$\ket{\psi}= a_0 \ket{0} + a_1 \ket{1}$ (with $|a_0|^2+|a_1|^2=1$)
can be regarded as an \emph{uncertain piece of information},
where the answer \emph{NO} has probability $|a_0|^2$, while the
answer\emph{YES} has probability $|a_1|^2$. The two
basis-elements $\ket{0}$ and $\ket{1}$ are taken as encoding the
classical bit-values $0$ and $1$, respectively. From a semantic
point of view, they can be also regarded as the classical
truth-values \emph{Falsity} and \emph{Truth}.

An $n$-qubit system (or \emph{$n$-quregister}) is represented by a
unit vector in the $n$-fold tensor product Hilbert space
$\otimes^n \C^2:=
\underbrace{\C^2\otimes\ldots\otimes\C^2}_{n-times}$ (where
$\otimes^1 \C^2:= \C^2$).
 We will use $x,y,\ldots$ as variables ranging over the set
$\{0,1\}$. At the same time, $\ket{x},\ket{y},\ldots$ will range
over the basis $\mathcal{B}^{(1)}$. Any factorized unit vector
$\ket{x_1}\otimes\ldots\otimes \ket{x_n}$ of the space $\otimes^n
\C^2$ will be called an $n$--\emph{configuration} (which can be
regarded as a quantum realization of a classical bit sequence of
length $n$). Instead of $\ket{x_1}\otimes\ldots\otimes \ket{x_n}$
we will simply write $\ket{x_1,\ldots,x_n}$. Recall that the
dimension of $\otimes^n\C^2$ is $2^n$, while the set of all
$n$--configurations
$\mathcal{B}^{(n)}=\{\ket{x_1,\ldots,x_n}:x_i\in \{0,1\}\}$ is an
orthonormal basis for the space $\otimes^n \C^2$. We will call
this set a \emph{computational basis} for the $n$--quregisters.
Since any string $x_1,\ldots,x_n$ represents a natural number
$j\in[0,2^n-1]$ (where $j=2^{n-1} x_1+2^{n-2} x_2+\ldots+x_n$),
any unit vector of $\otimes^n \C^2$ can be shortly expressed in
the following form: $\sum_{j=0}^{2^n-1} c_j \dket{j}$, where $c_j
\in \C$, $\dket{j}$ is the $n$-configuration corresponding to the
number $j$ and $\sum_{j=0}^{2^n-1} |c_j|^2=1$.

We will indicate by $\frak R(\otimes^n\C^2)$ the set of all
quregisters of $\otimes^n\C^2$, while $\frak R$ will represent the
set $\bigcup_{n=1}^{\infty}\frak R(\otimes^n\C^2)$. The set $\frak
R(\otimes^1\C^2)$ of all qubits will be shortly indicated by
$\frak Q$.

Consider now the two following sets of natural numbers:
$$ C_1^{(n)}:=\{i \,:\, \dket{i} = \ket{x_1,\ldots,x_n} \text{and } x_n = 1 \} $$
and
$$ C_0^{(n)}:=\{i \,:\, \dket{i} = \ket{x_1,\ldots,x_n} \text{and } x_n = 0 \}. $$

Let us refer to a generic unit vector of the space
$\otimes^n\C^2$:
$$ \ket{\psi} = \sum_{i=0}^{2^n-1} a_i \dket{i}. $$
We obtain:
$$ \ket{\psi}=\sum_{i\in C^{(n)}_0} a_i\dket{i}+\sum_{j\in C^{(n)}_1} a_j\dket{j}. $$

Let $P_1^{(n)}$ and $P^{(n)}_0$ be the projections onto the span
of $\parg{\dket{i}\,:\,i\in C^{(n)}_1}$ and
$\parg{\dket{i}\,:\,i\in C^{(n)}_0}$, respectively. Clearly,
$P_1^{(n)}+P_0^{(n)}=I^{(n)}$, where $I^{(n)}$ is the identity
operator of $\otimes^n\C^2$. Apparently, $P_1^{(n)}$ and
$P_0^{(n)}$ are density operators iff $n=1$. Let $k_n=
\dfrac{1}{2^{n-1}}$ be the normalization coefficient such that
$k_nP_1^{(n)}$ and $k_nP_0^{(n)}$ are density operators. From an
intuitive point of view, $k_nP_1^{(n)}$ can be regarded as a
privileged information corresponding to the \emph{Truth}, while
$k_nP_0^{(n)}$ corresponds to the \emph{Falsity}. In particular,
$P_1^{(1)}$ represents the bit $\ket{1}$, while $P_0^{(1)}$
represents the bit $\ket{0}$.
 Let
$\frak D(\otimes^n\C^2)$ be the set of all density operators of
$\otimes^n\C^2$ and let $\frak D:=\bigcup_{n=1}^{\infty}\frak
D(\otimes^n\C^2)$.

\begin{definition}(Qumix).\label{de:qumix} \nl
A \emph{qumix} is a density operator in $\frak D$.
\end{definition}

Needless to say, quregisters correspond to particular qumixs that
are \emph{pure states} (i.e. projections onto one-dimensional
closed subspaces of a given $\otimes^n\C^n$). For any quregister
$\ket{\psi}$, we will indicate by $P_{\ket{\psi}}$ the \emph{pure
density operator} represented by the projection onto the
one-dimensional subspace spanned by the vector $\ket{\psi}$. The
set of all pure density operators will be indicated by $\frak
D_{\frak R}$.

Recalling the Born rule, we can now define the \emph{probability-value} of any qumix.

\begin{definition}(Probability of a qumix). \label{de:probmix} \nl
For any qumix $\rho \in \frak D(\otimes^n\C^n)$:
$$ \Prob(\rho)=\Tr(P^{(n)}_1\rho). $$
\end{definition}

From an intuitive point of view, $ \Prob(\rho)$ represents the
probability that the information stocked by the qumix $\rho$ is
true. In the particular case where $\rho$ is a pure density
operator $P_{\ket{\psi}}$, determined by the qubit $\ket{\psi}=
a_0\ket{0}+a_1\ket{1}$, we obtain that $\Prob(\rho)= |a_1|^2$.

For any quregister $\ket{\psi}$, we will write $\Prob(\ket{\psi})$
instead of $\Prob(P_{\ket{\psi}})$.

\section{Quantum logical gates}

We will now introduce some examples of \emph{quantum logical
gates}. Generally, a quantum logical gate can be described as a
unitary operator, assuming arguments and values in a
product-Hilbert space $\otimes^n\C^2$. First of all we will study
the so called \emph{Petri-Toffoli gate} (\cite{Pe-67} and \cite{To-80}).
It will be expedient to start by analysing the simplest case,
where our Hilbert space has the form:
$$ \otimes^3\C^2=\C^2\otimes\C^2\otimes\C^2. $$
In such a case, the Petri-Toffoli gate transforms the vectors of
$\otimes^3\C^2$ into vectors of $\otimes^3\C^2$. In order to stress
that our operator is defined on the product space $\otimes^3\C^2$,
we will indicate it by $T^{(1,1,1)}$. Since we want to define a unitary
operator, it will be sufficient to determine its behaviour for the
elements of the basis, having the form
$\ket{x}\otimes\ket{y}\otimes\ket{z}$ (where $x,y,z \in \parg{0,1}$).

\begin{definition}(The Petri-Toffoli gate $T^{(1,1,1)}$). \label{de:toffoli} \nl
The \emph{Petri-Toffoli gate} $T^{(1,1,1)}$ is the linear
operator $T^{(1,1,1)}: \otimes^3\C^2 \to \otimes^3 \C^2$ that is
defined for any element $\ket{x}\otimes\ket{y}\otimes\ket{z}$ of
the basis as follows:
$$ T^{(1,1,1)}(\ket{x}\otimes\ket{y}\otimes\ket{z})=\ket{x}\otimes\ket{y}\otimes\ket{xy\oplus z}, $$
where $\oplus$ represents the sum modulo $2$.
\end{definition}

From an intuitive point of view, it seems quite natural to ``see''
the gate $T^{(1,1,1)}$ as a kind of self-reversible
``truth-table'' that transforms triples of \emph{zeros} and
\emph{ones} into triples of \emph{zeros } and \emph{ones}.
The ``table'' we obtain is the following:

\begin{align*}
\ket{0,0,0}\quad&\mapsto\quad\ket{0,0,0}\\
\ket{0,0,1}\quad&\mapsto\quad\ket{0,0,1}\\
\ket{0,1,0}\quad&\mapsto\quad\ket{0,1,0}\\
\ket{0,1,1}\quad&\mapsto\quad\ket{0,1,1}\\
\ket{1,0,0}\quad&\mapsto\quad\ket{1,0,0}\\
\ket{1,0,1}\quad&\mapsto\quad\ket{1,0,1}\\
\ket{1,1,0}\quad&\mapsto\quad\ket{1,1,1}\\
\ket{1,1,1}\quad&\mapsto\quad\ket{1,1,0}
\end{align*}

In the first six cases, $T^{(1,1,1)}$ behaves like the identity
operator; in the last two cases, instead, our gate transforms the
last element of the triple into the opposite element ($0$ is
transformed into $1$ and $1$ transformed into $0$).

One can easily show that $T^{(1,1,1)}$ has been well defined for
our aims: one is dealing with an operator that is not only linear
but also unitary.

By using $T^{(1,1,1)}$, we can introduce a convenient notion of
\emph{conjunction}. Our conjunction, which will be indicated by
$\QAnd$, is characterized as a function whose arguments are pairs
of vectors in $\C^2$ and whose values are vectors of the product
space $\otimes^3\C^2$.
\begin{definition}($\QAnd$). \label{de:and} \nl
For any $\ket{\varphi}\in\C^2$ and any
$\ket{\psi}\in \C^2$:
$$ \QAnd\part{\ket{\varphi},\ket{\psi}}:=T^{(1,1,1)}\part{\ket{\varphi}\otimes\ket{\psi}\otimes\ket{0}}. $$
\end{definition}
Clearly, the qubit $\ket{0}$ behaves here as an ``ancilla''.

Let us check that $\QAnd$ represents a good generalization of the
corresponding classical truth-function. For the arguments
$\ket{0}$ and $\ket{1}$ we will obtain the following
``truth-table'':
\begin{align*}
(\ket{0},\ket{0})\quad&\mapsto\quad
T^{(1,1,1)}(\ket{0}\otimes\ket{0}\otimes\ket{0})
=\ket{0}\otimes\ket{0}\otimes\ket{0} \\
(\ket{0},\ket{1})\quad&\mapsto\quad
T^{(1,1,1)}(\ket{0}\otimes\ket{1}\otimes\ket{0})
=\ket{0}\otimes\ket{1}\otimes\ket{0} \\
(\ket{1},\ket{0})\quad&\mapsto\quad
T^{(1,1,1)}(\ket{1}\otimes\ket{0}\otimes\ket{0})
=\ket{1}\otimes\ket{0}\otimes\ket{0} \\
(\ket{1},\ket{1})\quad&\mapsto\quad
T^{(1,1,1)}(\ket{1}\otimes\ket{1}\otimes\ket{0})
=\ket{1}\otimes\ket{1}\otimes\ket{1}
\end{align*}

One immediately realizes the difference with respect to the
classical case. The classical truth-table represents a typical
irreversible transformation:

\begin{align*}
(0,0)\quad&\mapsto\quad 0 \\
(0,1)\quad&\mapsto\quad 0 \\
(1,0)\quad&\mapsto\quad 0 \\
(1,1)\quad&\mapsto\quad 1
\end{align*}

The arguments of the function determine the value, but not the
other way around. As is well known, irreversibility generally
brings about dissipation of information. Mathematically, however,
any Boolean function $f:\{0,1\}^n\to\{0,1\}^m$ can be transformed
into a \emph{reversible} function $\hat{f}:\{0,1\}^{n}\times
\{0,1\}^{m}\to\{0,1\}^{n}\times \{0,1\}^{m}$ in the following way:
$$ \forall u \in \{0,1\}^n \: \forall v \in \{0,1\}^m: \hat{f}(u,v)=(u,v\oplus f(u)), $$
where $\oplus$ is the sum modulo 2 pointwise defined. The function that
is obtained by making reversible the irreversible classical
``and'' corresponds to the Petri-Toffoli gate. The classical
``and'' is then recovered by fixing the third input bit to 0.

Accordingly, the three arguments $(0,0)$, $(0,1)$, $(1,0)$ turn
out to correspond to three distinct values, represented by the
triples $(0,0,0)$, $(0,1,0)$, $(1,0,0)$. The price we have paid in
order to obtain a reversible situation is the increasing of the
complexity of our Hilbert space. The function $\QAnd$ associates
to pairs of arguments, belonging to the two-dimensional space
$\C^2$, values belonging to the space $\otimes^3\C^2$ (whose
dimension is $2^3$).

All this happens in the simplest situation, when one is only
dealing with elements of the basis (in other words, with precise
pieces of information). Let us examine the case where the function
$\QAnd$ is applied to arguments that are superpositions of the
basis-elements in the space $\C^2 $. Consider the following
\emph{qubit} pair:
$$ \ket{\psi}=a_0\ket{0}+a_1\ket{1}\,,\: \ket{\varphi}=b_0\ket{0}+b_1\ket{1}. $$
By applying the definitions of $\QAnd$ and of $T^{(1,1,1)}$, we obtain:
$$ \QAnd(\ket{\psi},\ket{\varphi})=a_1b_1\ket{1,1,1}+a_1b_0\ket{1,0,0}+ a_0b_1\ket{0,1,0}+ a_0b_0\ket{0,0,0}. $$

This result suggests a quite natural logical interpretation. The
four basis-elements that occur in our superposition-vector
correspond to the four cases of the truth-table for the classical
conjunction:
$$ (1,1,1), (1,0,0), (0,1,0), (0,0,0). $$
However here, unlike the classical situation, each case is accompanied
by a complex number, which represents a characteristic quantum
\emph{amplitude}. By applying the ``Born rule'' we will obtain
the following interpretation: $|a_1b_1|^2$ represents the
probability-value that both the \emph{qubit}-arguments are equal
to $\ket{1}$, and consequently their conjunction is $\ket{1}$.
Similarly in the other three cases.

So far we have considered a very special situation, characterized
by a Hilbert space having the form $\otimes^3\C^2$. However, our
procedure can be easily generalized. The Petri-Toffoli gate can be
defined in any Hilbert space having the form:

$$ (\otimes^n\C^2)\otimes(\otimes^m\C^2)\otimes\C^2(=\otimes^{n+m+1}\C^2). $$

\begin{definition}(The Petri-Toffoli gate $T^{(n,m,1)}$). \label{de:toffoligenerale} \nl
The \emph{Petri-Toffoli gate} $T^{(n,m,1)}$ is the linear operator
$$ T^{(n,m,1)}: (\otimes^n\C^2)\otimes(\otimes^m\C^2)\otimes\C^2\: \to \:
    (\otimes^n\C^2)\otimes(\otimes^m\C^2)\otimes\C^2, $$
that is defined for any element
$\ket{x_1,\ldots,x_n}\otimes\ket{y_1,\ldots,y_m}\otimes\ket{z}$ of
the computational basis of $\otimes^{n+m+1}\C^2$ as follows:
$$ T^{(n,m,1)}(\ket{x_1,\ldots,x_n}\otimes\ket{y_1,\ldots,y_m}\otimes\ket{z})
    =\ket{x_1,\ldots,x_n}\otimes\ket{y_1,\ldots,y_m}\otimes\ket{x_n y_m \oplus z},$$
where $\oplus$ represents the sum modulo $2$.
\end{definition}

On this basis one can immediately generalize our definition of
$\QAnd$.

\begin{definition}($\QAnd$). \nl
For any $\ket{\varphi}\in\otimes^n\C^2$ and any
$\ket{\psi}\in\otimes^m\C^2$:
$$ \QAnd\part{\ket{\varphi},\ket{\psi}}:
    =T^{(n,m,1)}\part{\ket{\varphi}\otimes\ket{\psi}\otimes\ket{0}}. $$
\end{definition}

How to deal in this context with the concept of negation? A
characteristic of quantum computation is the possibility of
defining a plurality of negation-operations: some of them
represent good generalizations of the classical negation. We will
first consider the operator $\tt Not^{(n)}$ that simply inverts
the value of the last element of any configuration of the space
$\otimes^n\C^2$. Thus, if $\ket{x_1,\ldots,x_n}$ is a vector of
the computational basis $\mathcal B^{(n)}$, the result of the
application of $\tt Not^{(n)}$ to $\ket{x_1,\ldots,x_n}$ will be
$\ket{x_1,\ldots,x_{n-1},1-x_n}$.

\begin{definition}($\QNot^{(n)}$). \label{de:notgenerale} \nl
The negation-gate is the linear operator $\QNot^{(n)}$ that is
defined for any element $\ket{x_1,\ldots,x_n}$ of the computational
basis of $\otimes^n\C^2$ as follows:
$$ \QNot^{(n)}(\ket{x_1,\ldots,x_n})= \ket{x_1,\ldots,x_{n-1},1-x_n}. $$
\end{definition}

One can immediately check that $\QNot^{(n)}$ represents a good
generalization of the classical truth-table. Consider the
basis-elements $\ket{0}$ and $\ket{1}$ of the space $\C^2$. In
such a case we obtain:
\begin{align*}
\QNot^{(1)}(\ket{1})=\ket{0}; \\
\QNot^{(1)}(\ket{0})=\ket{1}.
\end{align*}

The matrix corresponding to $\QNot^{(1)}$ is:

$$
\begin{pmatrix}
0 & 1 \cr
1 & 0 \cr
\end{pmatrix}
$$

Both the negation-gate and the Petri-Toffoli gate can be uniformly
defined on the set $\frak R$ of all quregisters in the expected
way:
\begin{align*}
&\QNot(\ket{\psi}) :=
    \QNot^{(n)}(\ket{\psi}),\hspace{4cm} \text{if} \; \ket{\psi} \in \otimes^n \C^2 ;\\
&T(\ket{\psi}, \ket{\varphi},\ket{\chi}) :=
    T^{(n,m,1)}(\ket{\psi}, \ket{\varphi}, \ket{\chi}), \\
&\hspace{3.53cm}\text{if} \; \ket{\psi} \in \otimes^n \C^2, \;
    \ket{\varphi} \in \otimes^m \C^2 \; \text{and} \; \ket{\chi} \in \C^2.
\end{align*}

Finally, how to introduce a reasonable disjunction? A gate $\QOr$
can be naturally defined in terms of $\QAnd$ and $\QNot$ via de Morgan.

\begin{definition}($\QOr$). \label{de:demorgan} \nl
For any quregisters $\ket{\varphi}$ and $\ket{\psi}$:
$$ \QOr(\ket{\varphi},\ket{\psi})=\QNot\part{\QAnd\part{\QNot(\ket{\varphi}),\QNot({\ket{\psi}})}}. $$
\end{definition}

At first sight, $\QAnd$ and $\QOr$ may look as
\emph{irreversible transformations}. However, it is important to
recall that, in this framework, $\QAnd(\ket{\psi},\ket{\varphi})$ should be regarded as a mere
metalinguistic abbreviation for
 $T(\ket{\psi},\ket{\varphi},\ket{0})$ (where $T$ is reversible).
Similarly $\QOr$.

 The quantum logical gates we have considered
so far are, in a sense, ``semiclassical''. A quantum logical
behaviour only emerges in the case where our gates are applied to
superpositions. When restricted to classical registers, our gates
turn out to behave as classical truth-functions. We will now
investigate \emph{genuine quantum gates} that may transform
classical registers into quregisters that are superpositions.

One of the most significant genuine quantum gates is the
\emph{square root of the negation}, which will be
indicated by $\QSNot$. As suggested by the name, the
characteristic property of the gate $\QSNot$ is the following: for
any quregister $\ket{\psi}$,
$$ \QSNot (\QSNot(\ket{\psi}))= \QNot( \ket{\psi}). $$

In other words: applying twice the square root of the negation
``means'' negating.

Interestingly enough, the gate $\QSNot$ has some
natural physical models (and implementations). As an example,
consider an idealized atom with a single electron and two energy
levels: a \emph{ground state} (identified with $\ket{0}$) and an
\emph{excited state} (identified with $\ket{1}$). By shining a
pulse of light of appropriate intensity, duration and wavelength,
it is possible to force the electron to change energy level. As a
consequence, the state (bit) $\ket{0}$ is transformed into the
state (bit) $\ket{1}$, and viceversa:
$$ \ket{0}\mapsto \ket{1}; \; \ket{1}\mapsto \ket{0}. $$

We have obtained a typical physical model for the gate $\QNot$.
Now, by using a light pulse of half the duration as the one
needed to perform the $\QNot$ operation, we effect a half-flip
between the two logical states. The state of the atom after the
half pulse is neither $\ket{0}$ nor $\ket{1}$, but rather a
superposition of both states. As observed by Deutsch, Ekert,
Lupacchini (\cite{DEL-00}):
\begin{quote}
Logicians are now entitled to propose a new logical operation
$\QSNot$. Why? Because a faithful physical model for it
exists in nature.
\end{quote}

The physical models of the gate $\QSNot$ naturally suggest
the following logical interpretation: $\QSNot$ represents a
kind of ``tentative negation''. By applying twice our ``attempt''
to negate, we obtain a full negation.

Interestingly enough, the gate $\QSNot$ seems to have also
some linguistic ``models''. For instance, consider the French
language. Put:
$$ \QSNot =\; \text{``ne''}=\; \text{``pas''}. $$
We obtain:
$$ \QSNot \QSNot= \; \text{``ne...pas''}\;=\QNot. $$
Needless to observe, our linguistic example is only a partial
model of the gate $\QSNot$. In French, neither the expression
``il ne pleut'' nor the expression ``il pleut pas'' are grammatically
correct sentences. And in the spoken language ``il pleut pas'' is
simply used as an abbreviation for the correct ``il ne pleut
pas''. In quantum computation, instead, for any quregister
$\ket{\psi}$, the vector $\QSNot(\ket{\psi})$ is a
quregister that is essentially different from the quregister $\QNot(\ket{\psi})$.

Let us now give the mathematical definition of $\QSNot$.

\begin{definition}(The square root of the negation). \label{de:radicenegazione} \nl
The square root of the negation on $\otimes^{n} \C^2$ is the
linear operator $\QSNot^{(n)}$ such that for every
element $\ket{x_1,\ldots,x_n}$ of the computational basis
$\mathcal{B}^{(n)}$:
$$ \QSNot^{(n)}(\ket{x_1,\ldots,x_n})
    =\ket{x_1,\ldots,x_{n-1}} \otimes (\dfrac{1+i}{2} \ket{x_n} + \dfrac{1-i}{2} \ket{1-x_n}) $$
(where $i$ is the imaginary unit).
\end{definition}
In other words,$\QSNot^{(n)}$ transforms the last
element $x_n$ of any configuration $\ket{x_1,\ldots,x_n}$ into the
element $\dfrac{1+i}{2} \ket{x_n} + \dfrac{1-i}{2} \ket{1-x_n}$. As a
consequence, for the two bits $\ket{0}$ and $\ket{1}$ (``living in
the space $\C^2$) we obtain:
$$ \QSNot^{(1)}(\ket{0})=\dfrac{1+i}{2} \ket{0} + \dfrac{1-i}{2} \ket{1}; $$
$$ \QSNot^{(1)}(\ket{1})=\dfrac{1-i}{2} \ket{0} + \dfrac{1+i}{2} \ket{1}. $$

One can easily show that $\QSNot^{(n)}$ is a unitary
operator, which satisfies the following condition:
$$ \text{for any } \ket{\psi} \in \otimes^{n} \C^2, \:\QSNot^{(n)}(\sqrt{\tt {Not}}^{(n)}(\ket{\psi}))
    =\QNot^{(n)} (\ket{\psi}). $$
In other words, applying twice the square root of
the negation means negating.

It turns out that the matrix associated to $\QSNot^{(1)}$
is
$$
\begin{pmatrix}
\dfrac{1+i}{2} & \dfrac{1-i}{2}\\[0.3cm]
\dfrac{1-i}{2} & \dfrac{1+i}{2}
\end{pmatrix}
$$

Like the negation, also the square root of the negation can be
uniformly defined on the set $\frak R $ of all quregisters:

\begin{align*}
\QSNot(\ket{\psi}) :=
    \QSNot^{(n)}(\ket{\psi}),\hspace{4cm} \text{if} \; \ket{\psi} \in \otimes^n \C^2 .\\
\end{align*}

As expected, the square root of the negation has no Boolean
counterpart.
\begin{lemma}
There is no function $f: \{0,1\} \to \{0,1\}$ such that for any
$x\in~\{0,1\}: \, f(f(x))= 1-x$.
\end{lemma}
\begin{proof}
Suppose, by contradiction, that such a function $f$ exists. Two
cases are possible: (i)\space $f(0)=0$; (ii)\space $f(0)=1$.
\par\noindent
(i)\space By hypothesis, $f(0)=0$. Thus, $1=f(f(0))=f(0)=0$,
contradiction.
\par\noindent
(ii)\space By hypothesis, $f(0)=1$. Thus, $1=f(f(0))=f(1)$. Hence,
$f(0)=f(1)$. Therefore, $1=f(f(0))=f(f(1))=0$, contradiction.
\end{proof}

Interestingly enough,$\QSNot$ does not even have any fuzzy
counterpart, represented by a continuous function (\cite{DCGLL-02}).
\begin{lemma}
There is no continuous function $f:[0,1] \to [0,1]$ such that for
any $x \in~[0,1]: \,f(f(x))=1-x$.
\end{lemma}
\begin{proof}
Suppose, by contradiction, that such a function $f$ exists. First,
we prove that $f(\frac{1}{2})=\frac{1}{2}$. By hypothesis,
$f(f(\frac{1}{2}))=1-\frac{1}{2}=\frac{1}{2}$. Hence,
$f(f(f(\frac{1}{2})))=f(\frac{1}{2})$. Thus,
$1-f(\frac{1}{2})=f(\frac{1}{2})$. Therefore,
$f(\frac{1}{2})=\frac{1}{2}$. Consider now $f(0)$. One can easily
show: $f(0)\not=0$ and $f(0)\not=1$. Clearly,
$f(0)\not=\frac{1}{2}$ since otherwise we would obtain
$1=f(f(0))=f(\frac{1}{2})=\frac{1}{2}$. Thus, only two cases are
possible: (i)\space $0<f(0)<\frac{1}{2}$; (ii)\space
$\frac{1}{2}<f(0)<1$.
\par\noindent
(i)\space By hypothesis, $0<f(0)<\frac{1}{2}<1=f(f(0))$.
Consequently, by continuity, $\exists x\in(0,f(0))$ such that
$\frac{1}{2}=f(x)$. Accordingly,
$\frac{1}{2}=f(\frac{1}{2})=f(f(x))=1-x$. Hence, $x=\frac{1}{2}$,
which contradicts $x<f(0)<\frac{1}{2}$.
\par\noindent
(ii)\space By hypothesis,
$f(\frac{1}{2})=\frac{1}{2}<f(0)<1=f(f(0))$. By continuity,
$\exists x\in(\frac{1}{2},f(0))$ such that $f(x)=f(0)$. Thus,
$1-x=f(f(x))=f(f(0))=1$. Hence, $x=0$, which contradicts
$x>\frac{1}{2}$.
\end{proof}

The gates considered so far can be naturally generalized to
qumixs. When our gates will be applied to density operators, we
will write: $\NOT$, $\SNOT$, $\AND$, $\OR$ (instead of $\QNot$, $\QSNot$, $\QAnd$, $\QOr$).

\begin{definition}(The negation). \label{de:negazionedensa} \nl
For any qumix $\rho\in\mathfrak D(\otimes^n\C^2)$,
$$ \NOT^{(n)}\rho=\QNot^{(n)}\rho \QNot^{(n)}. $$
\end{definition}

\begin{definition}(The square root of the negation). \label{de:radicedensa} \nl
For any qumix $\rho\in\mathfrak D(\otimes^n\C^2)$,
$$ \SNOT^{(n)}\rho=\QSNot^{(n)}\,\rho\QSNot^{(n)}\,^{\ast} $$
(where $\QSNot^{(n)}\,^\ast$ is the adjoint of
$\QSNot^{(n)}$).
\end{definition}

It is easy to see that for any $n\in\N^+$, both $\NOT^{(n)}(\rho)$
and $\SNOT^{(n)}(\rho)$ are qumixs of $\mathfrak
D(\otimes^n\C^2)$. Further: $\NOT^{(n)}\NOT^{(n)}=I^{(n)}$.

\begin{definition}(The conjunction). \label{de:congiunzionedensa} \nl
Let $\rho\in\mathfrak D(\otimes^n\C^2)$ and $\sigma\in\mathfrak
D(\otimes^m\C^2)$.
$$ \AND(\rho,\sigma)=T^{(n,m,1)}(\rho\otimes\sigma\otimes P^{(1)}_0) T^{(n,m,1)}. $$
\end{definition}

Like in the quregister-case, the gates $\NOT$, $\SNOT$, $\AND$,
$\OR$ can be uniformly defined on the set $\mathfrak D$ of all
qumixs.

The following theorem sums up some basic properties of our gates:

\begin{theorem}\label{th:guddernot}\nl
\begin{enumerate}
\item[(i)]$\SNOT\,\SNOT\,\rho= \NOT\rho;$
\item[(ii)]\space $\Prob(\NOT\,\rho)=1-\Prob(\rho)$;
\item[(iii)]\space $ \Prob(\SNOT\,\NOT\,\rho)=\Prob(\NOT\,\SNOT\,\rho)$;
\item[(iv)] \space $\Prob(\AND(\rho,\sigma))=\Prob(\rho)\Prob(\sigma);$
\item[(v)] $\Prob(\SNOT\,\AND(\rho,\sigma))=\frac{1}{2}.$
\end{enumerate}
\end{theorem}
\begin{proof}
\cite{Gu-02} and \cite{CDCGL-03}
\end{proof}

\section{Reversible and irreversible quantum computational structures}

An interesting feature of the qumix system is the following: any
real number $\lambda\in[0,1]\subset\R$ uniquely determines a qumix
$\rho^{(n)}_\lambda$ (for any $n\in\N^+$):

\begin{equation}
\rho^{(n)}_\lambda:= (1-\lambda) k_n P^{(n)}_0+\lambda k_n
P^{(n)}_1.
\end{equation}

Clearly, $\rho^{(n)}_\lambda\in\mathfrak D(\otimes^n\C^2)$. From
an intuitive point of view, $\rho^{(n)}_\lambda$ represents a
\emph{mixture of pieces of information} that might correspond to
the \emph{Truth} with probability $\lambda$.

From a physical point of view, $\rho^{(n)}_\lambda$ corresponds to
a particular preparation of the system such that the quantum
system might be in the state $k_n P^{(n)}_0$ with probability
$1-\lambda$ and in the state $k_n P^{(n)}_1$ with probability
$\lambda$. It is worthwhile recalling that the random polarized
states of the photon are represented by the density operator
$\rho^{(1)}_{1/2}=\um I^{(1)}$.

Two important properties of the qumix $\rho^{(n)}_\lambda$ are
described by the following lemma:

\begin{lemma}\label{le:rolambda} \nl
\begin{enumerate}
\item[(i)]\space $\forall n\in\N^+\,\forall\lambda\in[0,1]$: $\Prob(\rho^{(n)}_\lambda)=\lambda$;
\item[(ii)]\space $\Prob(\SNOT\rho^{(n)}_\lambda)=\um$.
\end{enumerate}
\end{lemma}
\begin{proof}
\cite{CDCGL-03}
\end{proof}

We will now introduce two interesting relations that can be
defined on the set of all qumixs. Both of them turn out to be a
preorder-relation. We will speak of \emph{weak} and of
\emph{strong preorder}, respectively.
\begin{definition}(Weak preorder). \label{de:preordinedebole} \nl
$$ \rho\le\sigma \text{ iff } \Prob(\rho)\le\Prob(\sigma). 
$$
\end{definition}

\begin{definition}(Strong preorder). \label{de:preordineforte} \nl
$\rho\preceq\sigma\text{ iff } \text{the following conditions hold}$:
\begin{enumerate}
\item[(i)]\space $\Prob(\rho)\le\Prob(\sigma)$; 
\item[(ii)]\space $\Prob(\SNOT\sigma)\le\Prob(\SNOT\rho)$. 
\end{enumerate}
\end{definition}

Clearly, $\rho\preceq\sigma$ implies $\rho\le\sigma$, but not the
other way around. One immediately shows that both $\le$ and
$\preceq$ are reflexive and transitive, but not antisymmetric.
Counterexamples can be easily found in $\mathfrak D(\C^2)$.

Consider now the following structure:
\begin{equation}\label{reversiblealgebra}
\part{\mathfrak D \,,\preceq \,,\AND\,,\NOT\,,\SNOT\,,P^{(1)}_0,P^{(1)}_1\,,\rho^{(1)}_{1/2}}.
\end{equation}
We will call such a structure \emph{the standard reversible
quantum computational structure} (shortly \emph{the
RQC-structure}).

In the following we will generally write $I$, $P_0$, $P_1$ and
$\rho_{1/2}$ instead of $I^{(1)}$, $P_0^{(1)}$,$P_1^{(1)}$,
$\rho^{(1)}_{1/2}$. From an intuitive point of view, $P_0$, $P_1$
and $\rho_{1/2}$ represent privileged pieces of information that
are \emph{true}, \emph{false}, \emph{indeterminate}, respectively.
Generally, our qumixs fail to satisfy \emph{Duns Scotus law}:
$P_0$ and $P_1$ are not the \emph{minimum} and the \emph{maximum}
element of the RQC-structure. Hence, in this situation, it is
interesting to isolate the elements that have a \emph{Scotian}
behaviour.

\begin{definition}(Down and up scotian qumixs). \label{de:scotian} \nl
Let $\rho$ be a qumix of $\mathfrak D$.
\begin{enumerate}
\item[(i)]\space $\rho$ is \emph{down Scotian} iff $\,P_0 \preceq \rho$; 
\item[(ii)]\space $\,\rho$ is \emph{up Scotian} iff $\rho\preceq P_1$; 
\item[(iii)]\space $\rho$ is \emph{Scotian} iff $\rho$ is both down and up Scotian. 
\end{enumerate}
\end{definition}

\begin{lemma}\label{le:scotiano1} \nl
\begin{enumerate}
\item[(i)]\space $\rho\preceq\SNOT\, P_1\,$ iff $\,\Prob(\rho)\le\um$;
\item[(ii)]\space $\SNOT\,P_0\preceq\rho\,$ iff $\,\Prob(\rho)\ge \um$.
\end{enumerate}
\end{lemma}
\begin{proof}
\cite{CDCGL-03}
\end{proof}

\begin{theorem}\label{th:scotiano2} \nl
\begin{enumerate}
\item[(i)]\space $\rho$ is down Scotian iff $\,\Prob(\SNOT\rho)\le\um\,$ iff $\,\SNOT\rho\preceq\SNOT\,P_1$;
\item[(ii)]\space $\rho$ is up Scotian iff $\,\um \le \Prob(\SNOT\rho)\,$ iff $\,\SNOT P_0 \preceq \SNOT\rho$;
\item[(iii)]\space $\rho\,$ is Scotian iff $\,\Prob(\SNOT\rho)=\um$;
\item[(iv)]\space $\forall n\in\N^+$: $k_nP^{(n)}_0, k_nP^{(n)}_1, \rho^{(n)}_{1/2}\,$ are Scotian;
\item[(v)]\space For any $\in\N^+$, the set $\mathfrak D(\otimes^n\C^2)$ contains uncountably many
    Scotian density operators.
\end{enumerate}
\end{theorem}
\begin{proof}
\cite{CDCGL-03}
\end{proof}

The gates we have considered so far represent typical
\emph{reversible} logical operations. From a logical point of
view, it might be interesting to consider also some
\emph{irreversible} operations. An important example is
represented by a \Lu-like disjunction.
\begin{definition}(The \Lu \ disjunction). \nl
Let $\tau\in\mathfrak D(\otimes^n\C^2)$ and $\sigma\in\mathfrak D(\otimes^m\C^2)$.
$$ \tau\oplus\sigma:=\rho^{(1)}_{\Prob(\tau)\oplus\Prob(\sigma)}, $$
where $\oplus$ in $\Prob(\tau)\oplus\Prob(\sigma)$ is the \Lu \ ``truncated sum''
defined on the real interval $[0,1]$ (i.e.
$\Prob(\tau)\oplus\Prob(\sigma)=\min\parg{1,\Prob(\tau)+\Prob(\sigma)}$)\,(\cite{Za-34}).
\end{definition}

The following lemmas sum up some basic properties of the \Lu \
disjunction:

\begin{lemma} \label{le:lukone} \nl
\begin{enumerate}
\item[(i)]\space
$$
\tau\oplus\sigma =
\begin{cases}
    \rho^{(1)}_{\Prob(\tau)+\Prob(\sigma)},
    &\text{if }\Prob(\tau)+\Prob(\sigma)\le 1;\\[0.3cm]
    P^{(1)}_1, &\text{otherwise};
\end{cases}
$$
\item[(ii)]\space $\Prob(\tau\oplus\sigma)=\Prob(\tau)\oplus\Prob(\sigma)$;
\item[(iii)]\space $\Prob(\SNOT(\tau\oplus\sigma))=\um$.
\end{enumerate}
\end{lemma}
\begin{proof}
\cite{CDCGL-03}
\end{proof}

\begin{lemma}\label{le:solitone}
Let $\rho\in\mathfrak D(\otimes^n\C^2)$.
\begin{enumerate}
\item[(i)]\space $\forall n\in\N^+$: $\rho\oplus k_nP^{(n)}_1=P^{(1)}_1$;
\item[(ii)]\space $\forall n\in\N^+$: $\rho\oplus k_nP^{(n)}_0=\rho^{(1)}_{\Prob(\rho)}$;
\item[(iii)]\space $\rho\oplus\NOT\rho=P^{(1)}_1$.
\end{enumerate}
\end{lemma}
\begin{proof}
Straightforward.
\end{proof}

From Lemma \ref{le:solitone} it follows that $\Prob(\rho\oplus
k_nP^{(n)}_1)=1$, $\Prob(\rho\oplus k_nP^{(n)}_0)=\Prob(\rho)$ and
$\Prob(\rho\oplus\NOT\rho)=1$.

The preorder $\preceq$ permits us to define on the set of all
qumixs an equivalence relation in the expected way.

\begin{definition}(The strong equivalence relation). \label{de:congruenzaforte} \nl
$$ \rho\approxeq\sigma \text{ iff } \rho\preceq\sigma \text{ and } \sigma\preceq\rho. $$
\end{definition}

Clearly, $\approxeq$ is an equivalence relation. Let
$$ [\mathfrak D]_\approxeq:=\parg{[\rho]_\approxeq\,:\,\rho\in\mathfrak D}. $$
We will omit $\approxeq$ in $[\rho]_\approxeq$ if no confusion is possible.

Unlike the qumixs (which are only preordered by $\preceq$), the
equivalence-classes of $[\mathfrak D]_\approxeq$ can be partially
ordered in a natural way.

\begin{definition}
$$ [\rho]\preceq[\sigma] \text{ iff } \rho\preceq\sigma. $$
\end{definition}
The relation $\preceq$ (which is well defined) is a partial order.
\begin{lemma}\label{le:unmezzo}\nl
\begin{enumerate}
\item[(i)]\space $\forall n\in\N^+$: $\parq{P_1}=\parq{k_nP^{(n)}_1}$;
\item[(ii)]\space \space $\forall n\in\N^+$: $\parq{P_0}=\parq{k_nP^{(n)}_0}$;
\item[(iii)]\space $\forall n\in\N^+\,\forall\lambda\in[0,1]$:$\,\parq{\rho^{(1)}_\lambda}=\parq{\rho^{(n)}_\lambda}$.
\end{enumerate}
\end{lemma}
\begin{proof}
\cite{CDCGL-03}
\end{proof}

On this basis, one can naturally define on the set $[\mathfrak
D]_\approxeq$ a conjunction, a negation, the square root of the
negation, a \Lu \ disjunction:
\begin{definition} \label{de:operantiforti} \nl
Let $\rho\in\mathfrak D(\otimes^n\C^2)$ and $\sigma\in\mathfrak
D(\otimes^m\C^2)$.
\begin{enumerate}
\item[(i)]\space $[\rho]\AND [\sigma]=[\AND(\rho,\sigma)]$;
\item[(ii)]\space $\NOT[\rho]=[\NOT\rho]$;
\item[(iii)]\space $\SNOT[\rho]=[\SNOT\rho]$;
\item[(iv)]\space $[\rho]\oplus[\sigma]=[\rho\oplus\sigma]$.
\end{enumerate}
\end{definition}

\begin{lemma}\label{le:congruenteforte}
The operations of Definition \ref{de:operantiforti} are well defined.
\end{lemma}
\begin{proof}
\cite{CDCGL-03}
\end{proof}

\begin{lemma}\label{le:semigruppale}\nl
\begin{enumerate}
\item[(i)]\space  The operation $\AND$ is associative and commutative;
\item[(ii)]\space The operation $\oplus$ is associative and commutative;
\item[(iii)]\space  $\NOT\,\NOT[\rho]=[\rho]$;
\item[(iv)]\space $\SNOT\,\SNOT[\rho]=\NOT[\rho]$;
\item[(v)]\space $\SNOT\,\NOT[\rho]=\NOT\SNOT[\rho]$.
\end{enumerate}
\end{lemma}
\begin{proof}
Straightforward.
\end{proof}

Consider now the structure
\begin{equation}\label{standardalgebra}
\part{[\mathfrak D]_\approxeq\,,\AND \,,\oplus\,,\NOT\,,\SNOT\,,[P_0]_\approxeq,[P_1]_\approxeq\,,[\rho_{1/2}]}.
\end{equation}

We will call such a structure \emph{the standard irreversible
quantum computational algebra} (shortly \emph{the IQC-algebra}).

As happens in the case of $\preceq$, also the weak preorder $\le$
permits us to define an equivalence relation, which will be called
\emph{weak equivalence relation}.

\begin{definition}(Weak equivalence relation). \label{de:congurenzadebole} \nl
$$ \rho\equiv\sigma \text{ iff } \rho\le\sigma \text{ and } \sigma\le\rho. $$
\end{definition}
Clearly, $\equiv$ is an equivalence relation. Let
$$ [\mathfrak D]_\equiv:=\parg{[\rho]_\equiv\,:\,\rho\in\mathfrak D}. $$

Also $[\mathfrak D]_\equiv$ can be partially ordered in a natural
way.

\begin{definition}
$$ [\rho]_{\equiv}\le[\sigma]_{\equiv} \text{ iff } \rho\le\sigma. $$
\end{definition}
One can easily show that the relation $\le$ (which is well
defined) is a partial order.

A conjunction, a \Lu \ disjunction, a negation (but not the square
root of the negation!) can be naturally defined on $[\mathfrak
D]_\equiv$.

\begin{definition} \label{de:operantideboli} \nl
Let $\rho\in\mathfrak D(\otimes^n\C^2)$ and $\sigma\in\mathfrak
D(\otimes^m\C^2)$.
\begin{enumerate}
\item[(i)]\space $[\rho]_\equiv\AND[\sigma]_\equiv=[\AND(\rho,\sigma)]_\equiv$;
\item[(ii)]\space $\NOT[\rho]_\equiv=[\NOT\rho]_\equiv$;
\item[(iii)]\space $[\rho]_\equiv\oplus[\sigma]_\equiv=[\rho\oplus\sigma]_\equiv$.
\end{enumerate}
\end{definition}

\begin{lemma}\label{le:congruentedebole}
The operations of Definition \ref{de:operantideboli} are well defined.
\end{lemma}
\begin{proof}
\cite{CDCGL-03}
\end{proof}

Unlike $\approxeq$, the relation $\equiv$ is not a congruence with
respect to $\SNOT$. In fact, the following situation is possible:
$[\rho]_\equiv=[\sigma]_\equiv$ and
$[\SNOT\,\rho]_\equiv\not=[\SNOT\,\sigma]_\equiv$. Consider for
example the following unit vectors of $\C^2$:
$\ket{\psi}:=\dfrac{\sqrt{2}}{2}\ket{0}+\dfrac{\sqrt{2}}{2}\ket{1}$
and $\ket{\varphi}:=\dfrac{\sqrt{2}}{2}\ket{0}+\dfrac{1+i}{2}\ket{1}$.

Let $P_{\ket{\psi}}$ and $P_{\ket{\varphi}}$ be the projections
onto the unidimensional spaces spanned by $\ket{\psi}$ and
$\ket{\varphi}$, respectively. It turns out that
$\Prob(P_{\ket{\psi}})=\Prob(P_{\ket{\varphi}})=\um$. Accordingly,
$[P_{\ket{\psi}}]_\equiv=[P_{\ket{\varphi}}]_\equiv$. However,
$\Prob(\SNOT\,P_{\ket{\psi}}) =\um$ and
$\Prob(\SNOT\,P_{\ket{\varphi}})=\dfrac{1}{2} -
\dfrac{\sqrt{2}}{4}\approx 0.146447$. Consequently,
$[P_{\ket{\psi}}]_\approxeq\not=[P_{\ket{\varphi}}]_\approxeq$.

An interesting relation between the weak and the strong preorder
is described by the following theorem.

\begin{theorem}\label{th:debolepreordine}
For any $\rho,\sigma\in\mathfrak D$:
$$ [\rho]_\equiv\le[\sigma]_\equiv \text{ iff } [\rho]_\approxeq \,\AND\,[P_1]_\approxeq \preceq
    [\sigma]_\approxeq\,\AND\,[P_1]_\approxeq. $$
\end{theorem}
\begin{proof}
\cite{CDCGL-03}
\end{proof}

\section{The Poincar\'{e} quantum computational structures}

We will now restrict our analysis to the qumixs living in the
two-dimensional space $\C^2$. As is well known, every density
operator of $\mathfrak D(\C^2)$ has the following matrix
representation:

\begin{equation}
\um\part{I+r_1X+r_2Y+r_3Z},
\end{equation}
where $r_1,r_2,r_3$ are real numbers such that
$r_1^2+r_2^2+r_3^2\le 1$ and $X,Y,Z$ are the Pauli matrices:

$$
X=
\begin{pmatrix}
    0 & 1 \\
    1 & 0
\end{pmatrix}
\qquad Y=
\begin{pmatrix}
    0 & -i \\
    i & 0
\end{pmatrix}
\qquad Z=
\begin{pmatrix}
    1 & 0 \\
    0 & -1
\end{pmatrix}.
$$

It turns out that a density operator $\um\part{I+r_1X+r_2Y+r_3Z}$
is pure iff $r_1^2+r_2^2+r_3^2=1$. Consequently,
\begin{itemize}
\item Pure density operators are in $1:1$ correspondence with
    the points of the surface of the Poincar\'{e} sphere;
\item Proper mixtures are in $1:1$ correspondence with
    the inner points of the Poincar\'{e} sphere.
\end{itemize}
Let $\rho$ be a density operator of $\mathfrak D(\C^2)$. We will
denote by $\bar{\rho}$ the point of the Poincar\'{e} sphere that is
univocally associated to $\rho$.

Let $(r_1,r_2,r_3)$ be a point of the Poincar\'{e} sphere. We will
denote by $\widehat{(r_1,r_2,r_3)}$ the density operator
univocally associated to $(r_1,r_2,r_3)$.

\begin{lemma}\label{le:soloindue}
Let $\rho\in\mathfrak D(\C^2)$ such that
$\bar{\rho}=(r_1,r_2,r_3)$. The following conditions hold:
\begin{enumerate}
\item[(i)]\space $\Prob(\rho)=\dfrac{1-r_3}{2}$ and  $\Prob(\SNOT\,\rho)=\dfrac{1-r_2}{2}$;
\item[(ii)]\space $0 < \Prob(\rho)< 1$ and  $0 < \Prob(\SNOT\rho)< 1$, whenever  $\rho$ is a proper mixture.
\end{enumerate}
\end{lemma}
\begin{proof}\nl

\begin{enumerate}
\item[(i)] \space Easy computation;
\item[(ii)] \space Since proper mixtures are in 1:1 correspondence with inner points
 of the Poincar\'{e} sphere, we have: $r^2_1+r^2_2+r^2_3<1$. Hence: $r^2_2,r^2_3<1$ and
 $-1 < r_2,r_3 <1$. Consequently: $0 < \Prob(\rho)= \dfrac{1-r_3}{2}< 1$ and
 $0 < \Prob(\SNOT\rho)= \dfrac{1-r_2}{2}< 1$.
\end{enumerate}

\end{proof}

An irreversible conjunction can be now naturally defined on the
set of all qumixs of $\mathfrak D(\C^2)$.
\begin{definition}(The irreversible conjunction). \label{de:andirreversibile} \nl
Let $\tau,\sigma\in\mathfrak D(\C^2)$.
\begin{equation}
\IAND(\tau,\sigma):=\rho^{(1)}_{\Prob(\tau)\Prob(\sigma)}
\end{equation}
\end{definition}

Interestingly enough, the density operator $\IAND(\tau,\sigma)$
can be described in terms of the \emph{partial trace}. Suppose we
have a compound physical system consisting of three subsystems,
and let
$$ \mathcal H=(\otimes^n\C^2)\otimes (\otimes^m\C^2)\otimes (\otimes^r\C^2) $$
be the Hilbert space associated to our system. Then, for any
density operator $\rho$ of $\mathcal H$, there is a unique density
operator $\Tr_{1,2}(\rho)$ that represents the \emph{partial
trace} of $\rho$ on the space $\otimes^r\C^2$ (associated to the
third subsystem). The two operators $\rho$ and $\Tr_{1,2}(\rho)$
are statistically equivalent with respect to the third subsystem.
In other words, for any self-adjoint operator $A^{(r)}$ of
$\otimes^r\C^2$:
$$ \Tr(\Tr_{1,2}(\rho)\, A^{(r)})= \Tr(\rho\, (I^{(n)}\otimes I^{(m)} \otimes A^{(r)})). $$

The density operator $\Tr_{1,2}(\rho)$, obtained by ``tracing
out'' the first and the second subsystem, is also called the
\emph{reduced state} of $\rho$ on the third subsystem.

One can prove that:
$$ \IAND(\tau,\sigma)=\Tr_{1,2}(\AND(\tau,\sigma)). $$

In other words, $\IAND(\tau,\sigma)$ represents the reduced state
of $\AND(\tau,\sigma)$ on the third subsystem.

An interesting situation arises when both $\tau$ and $\sigma$ are
pure states. For instance, suppose that:
$$ \tau = P_{\ket{\psi}}\:\text{and}\:\sigma = P_{\ket{\varphi}}, $$
where $\ket{\psi}$ and $\ket{\varphi}$ are proper qubits. Then,
$$ \AND(\tau,\sigma) = P_{T^{(1,1,1)}(\ket{\psi}\otimes \ket{\varphi}\otimes \ket{0})}, $$
which is a pure state. At the same time, we have:
$$\IAND(\tau,\sigma) = \Tr_{1,2}(P_{T^{(1,1,1)}(\ket{\psi}\otimes \ket{\varphi}\otimes \ket{0})}), $$
which is a proper mixture. Apparently, when considering only the
properties of the third subsystem, we loose some information.
As a consequence, we obtain a final state that does not represent
a maximal knowledge. As is well known, situations where the state
of a compound system represents a maximal knowledge, while the
states of the subsystems are proper mixtures, play an important
role in the framework of entanglement-phenomena.

\begin{lemma}\label{le:iand} \nl
\begin{enumerate}
\item[(i)]\space $\IAND$ is associative and commutative;
\item[(ii)]\space $\IAND(\rho,P_0)=P_0$;
\item[(iii)]\space  $\IAND(\rho,P_1)=\rho_{\Prob(\rho)}$;
\item[(iv)]\space $\Prob(\IAND(\rho,\sigma))=\Prob(\rho)\Prob(\sigma)$;
\item[(v)]\space $\Prob(\SNOT\,\IAND(\rho,\sigma))=\um$.
\end{enumerate}
\end{lemma}
\begin{proof}
Easy.
\end{proof}

Consider now the structure
\begin{equation}\label{bigstandardalgebra}
\part{\mathfrak D(\C^2)\,,\IAND \,,\oplus\,,\NOT\,,\SNOT\,,P_0,P_1\,,\rho_{1/2}}.
\end{equation}

We will call such a structure \emph{the Poincar\'{e} irreversible
quantum computational algebra} (shortly \emph{the Poincar\'{e}
IQC-algebra}).

We can refer to the relation $\upharpoonright\approxeq$,
representing the restriction of $\approxeq$ to $\mathfrak
D(\C^2)$. For any $\rho\in\mathfrak D(\C^2)$, let
\begin{equation}
[\rho]_{\upharpoonright\approxeq}:=\parg{\sigma\in\mathfrak
D(\C^2)\,:\,\rho\approxeq\sigma}.
\end{equation}

Further define
\begin{equation}
[\mathfrak
D(\C^2)]_{\upharpoonright\approxeq}:=\parg{[\rho]_{\upharpoonright\approxeq}
\,:\,\rho\in\mathfrak D(\C^2)}.
\end{equation}

The operations $\IAND\,,\oplus\,,\NOT\,,\SNOT$ and the relation
$\preceq$ can be defined on $\mathfrak
[D(\C^2)]_{\upharpoonright\approxeq}$ in the expected way.

Consider now the quotient-structure
$$\part{[\mathfrak D(\C^2)]_{\upharpoonright\approxeq}\,,\IAND\,,\oplus\,,\NOT\,,\SNOT\,,
    [P_0]_{\upharpoonright\approxeq}\,,[P_1]_{\upharpoonright\approxeq}
    \,,[\rho_{1/2}]_{\upharpoonright\approxeq}}. $$
We will call such a structure \emph{the contracted Poincar\'{e}
irreversible quantum computational algebra} (shortly
\emph{the contracted Poincar\'{e} IQC-algebra}).

\begin{theorem}\label{th:immersa}
The contracted Poincar\'{e} IQC-algebra is isomorphic to the
IQC-algebra, via the map $g:\,[\mathfrak
D(\C^2)]_{\upharpoonright\approxeq}\to [\mathfrak D]_\approxeq$
such that $\forall\rho\in\mathfrak D(\C^2)$:
\begin{equation}
g([\rho]_{\upharpoonright\approxeq})=[\rho]_\approxeq.
\end{equation}
Further, for any $\rho\,,\sigma\in\mathfrak D(\C^2)$:
$\,[\rho]_{\upharpoonright\approxeq} \preceq
\,[\sigma]_{\upharpoonright\approxeq}\,$ iff
$g(\,[\rho]_{\upharpoonright\approxeq}) \preceq
\,g([\sigma]_{\upharpoonright\approxeq})$.
\end{theorem}
\begin{proof}
\cite{CDCGL-03}
\end{proof}

One can prove that any density operator $\rho$ in $\frak D(\C^2)$
is associated to a qubit $\ket{\psi_{\rho}}$ that is ``statistically
equivalent'' to $\rho$. In a sense, $\ket{\psi_{\rho}}$ represents a
``purification'' of $\rho$.

\begin{lemma} \label{le:purissimo}\nl
For any $\rho \in \frak D(\C^2)$ such that
$\bar{\rho}=(r_1,r_2,r_3)$, there exists a qubit
$\ket{\psi_{\rho}}$ that satisfies the following conditions:
\begin{enumerate}
\item[(i)] \space $\Prob(\rho) = \Prob(\ket{\psi_{\rho}})$;
\item[(ii)] \space $\Prob(\SNOT\rho) = \Prob(\QSNot(\ket{\psi_{\rho}}))$.
\end{enumerate}
\end{lemma}
\begin{proof}
Let $\rho \in \frak D(\C^2)$ such that $\bar{\rho}=(r_1,r_2,r_3)$.
Consider the vector
$$ \ket{\psi_{\rho}}=\frac{\sqrt{1-r^2_2-r^2_3}-ir_2}{\sqrt{2(1-r_3)}}\ket{0}+ \sqrt{\frac{1-r_3}{2}}\ket{1}, $$
which turns out to be a qubit. An easy computation shows that
$$ \Prob(\ket{\psi_{\rho}})=\frac{1-r_3}{2}\quad \text{and}\quad \Prob(\QSNot\ket{\psi_{\rho}})=\frac{1-r_2}{2}. $$
Thus by Lemma \ref{le:soloindue} (i), we can conclude that
$$ \Prob(\ket{\psi_{\rho}})= \Prob(\rho) \quad \text{and} \quad \Prob(\QSNot(\ket{\psi_{\rho}}))= \Prob(\SNOT \rho). $$
\end{proof}

As an interesting application of Lemma \ref{le:purissimo} consider
a density operator whose form is: $\rho_{\lambda}= (1-\lambda)P_0
+ \lambda P_1$. Then, by Lemma \ref{le:purissimo}, there exists a
qubit $\ket{\psi_{\rho_{\lambda}}}$ such that
$\Prob(\ket{\psi_{\rho_{\lambda}}})= \lambda$. It turns out that
$$ \ket{\psi_{\rho_{\lambda}}}= \sqrt{1-\lambda}\ket{0}+\sqrt{\lambda}\ket{1}. $$

\begin{theorem}\label{th:rompiscatole}
Let $f: \frak D^n \to \frak D(\C^2)$. Consider the set $\frak Q$
of all qubits. Then, there exists a map
$$f_{\frak Q}: \frak Q^n \to \frak Q $$
such that for any qubits
$\ket{\psi_1},\ldots,\ket{\psi_n}$ the following conditions hold:
\begin{enumerate}
\item[(i)] \space $\Prob(f_{\frak Q}(\ket{\psi_1},\ldots,\ket{\psi_n}))=
\Prob(f(P_{\ket{\psi_1}},\ldots,P_{\ket{\psi_n}}));$
\item[(ii)] \space $\Prob( \QSNot(f_{\frak Q}(\ket{\psi_1},\ldots,\ket{\psi_n})))=
\Prob(\SNOT f(P_{\ket{\psi_1}},\ldots,P_{\ket{\psi_n}})).$
\end{enumerate}
\end{theorem}
\begin{proof}
Let $\ket{\psi_1},\ldots,\ket{\psi_n} \in \frak{Q}$.
Then $P_{\ket{\psi_1}},\ldots,P_{\ket{\psi_n}} \in \frak{D}$ and
$f(P_{\ket{\psi_1}},\ldots,P_{\ket{\psi_n}}) \in \frak{D}(\C^2)$.
By lemma \ref{le:purissimo}, there exists a qubit
$\ket{\psi_{f(P_{\ket{\psi_1}},\ldots,P_{\ket{\psi_n}})}}$
such that $\Prob(f(P_{\ket{\psi_1}},\ldots,P_{\ket{\psi_n}}))
    =\Prob(\ket{\psi_{f(P_{\ket{\psi_1}},\ldots,P_{\ket{\psi_n}})}})$
and $\Prob(\SNOT f(P_{\ket{\psi_1}},\ldots,P_{\ket{\psi_n}}))
    =\Prob(\QSNot(\ket{\psi_{f(P_{\ket{\psi_1}},\ldots,P_{\ket{\psi_n}})}}))$.
Thus, we can put
$f_{\frak Q}(\ket{\psi_1},\ldots,\ket{\psi_n}):=\ket{\psi_{f(P_{\ket{\psi_1}},\ldots,P_{\ket{\psi_n}})}}$.
\end{proof}

As a significant application of Theorem \ref{th:rompiscatole}, we
obtain that a \Lu \ disjunction $\oplus_{\frak Q}$ and an
irreversible conjunction ${\tt IAnd}_{\frak Q}$ can be naturally
defined for any qubits $\ket{\varphi}=a_0\ket{0}+a_1\ket{1}$ and
$\ket{\chi}=b_0\ket{0}+b_1\ket{1}$:
$$ \ket{\varphi} \oplus_{\frak Q} \ket{\chi}:=
\begin{cases}
    \sqrt{1-|a_1|^2-|b_1|^2}\ket{0}+\sqrt{|a_1|^2+|b_1|^2}\ket{1},&\text{if } |a_1|^2+|b_1|^2 \leq 1;\\[0.3cm]
    \ket{1}, &\text{otherwise};
\end{cases}
$$
$$ {\tt IAnd}_{\frak Q}(\ket{\varphi},\ket{\chi}):=\sqrt{1-|a_1 b_1|^2}\ket{0}+|a_1 b_1|\ket{1}. $$

From an intuitive point of view, it is interesting to compare
${\tt IAnd}_{\frak Q}(\ket{\varphi},\ket{\chi})$ with ${\tt
IAND}(P_{\ket{\varphi}},P_{\ket{\chi}})$ and with $\QAnd(\ket{\varphi},\ket{\chi})$.
As we already know, $\QAnd(\ket{\varphi},\ket{\chi})$ represents a pure state of a
compound physical system (living in the space $\otimes^3\C^2$).
Hence, one is dealing with a \emph{maximal knowledge}, that also
includes a maximal knowledge about the component systems
(described by the pure states $\ket{\varphi}$ and $\ket{\chi}$,
respectively). Further, the transformation
$(\ket{\varphi},\ket{\chi}) \mapsto \QAnd(\ket{\varphi},\ket{\chi})$ is reversible. The state ${\tt
IAND}(P_{\ket{\varphi}},P_{\ket{\chi}})$, instead, is generally a
proper mixture: a non-maximal knowledge about a (non-decomposed)
system, representing the output of a computation, where the
original information about the component systems (the inputs) has
been lost. The transformation $(P_{\ket{\varphi}},P_{\ket{\chi}})
\mapsto {\tt IAnd}(P_{\ket{\varphi}},P_{\ket{\chi}})$ is typically
irreversible. The state ${\tt IAnd}_{\frak
Q}(\ket{\varphi},\ket{\chi})$ represents a ``purification'' of
${\tt IAND}(P_{\ket{\varphi}},P_{\ket{\chi}})$: one is dealing
with a maximal knowledge about the output, that does not preserve
the original information about the inputs.

\section{Quantum computational logics}

The quantum computational structures we have investigated suggest
a natural semantics, based on the following intuitive idea: any
sentence $\alpha$ of the language is interpreted as a convenient
qumix, that generally depends on the logical form of $\alpha$; at
the same time, the logical connectives are interpreted as
operations that either are \emph{gates} or can be conveniently
simulated by gates. We will consider a \emph{minimal (sentential)
quantum computational language} $\mathcal{L}$ that contains a
privileged atomic sentence $\mathbf f$ (whose intended
interpretation is the truth-value \emph{Falsity}) and the
following primitive connectives: the \emph{negation} ($\lnot$),
the \emph{square root of the negation } ($\sqrt{\lnot}$), the
\emph{conjunction} ($\land$). Let $\FM$ be the set of all
sentences of $\mathcal{L}$. We will use the following
metavariables: $\mathbf q,\mathbf r \ldots$ for
atomic sentences and $\alpha,\beta,\ldots$ for sentences. The
connective disjunction ($\lor$) is supposed to be defined via de
Morgan ($\alpha \lor \beta := \lnot(\lnot \alpha \land \lnot
\beta)$), while the privileged sentence $\mathbf t$ representing
the \emph{Truth} is defined as the negation of $\mathbf f$
($\mathbf t:= \mathbf \lnot \mathbf f$). This minimal quantum
computational language can be extended to richer languages
containing other primitive connectives (for instance, a connective
corresponding to the \Lu \ irreversible disjunction $\oplus$) that
we will not consider here.

We will first introduce the notion of \emph{reversible quantum
computational model} (shortly, \emph{RQC-model}).
\begin{definition}(RQC-model). \nl
A \emph{RQC-model} of $\mathcal{L}$ is a function
$\Qum: Form^\mathcal{L} \to \mathfrak D$
(which associates to any sentence $\alpha$ of the language a qumix): \\
$\Qum(\alpha) :=
\begin{cases}
    \text{a density operator of $\mathfrak D(\C^2)$} & \text{if}\: \alpha\: \text{ is an atomic sentence}; \\
    P_0 & \text{if}\: \alpha=\mathbf f;\\
    \NOT \, \Qum(\beta) & \text{if} \: \alpha = \lnot \beta; \\
    \SNOT \,\Qum(\beta) & \text{if} \: \alpha = \sqrt{\lnot} \beta; \\
    \AND(\Qum(\beta),\Qum(\gamma)) & \text{if} \: \alpha = \beta \land \gamma.
\end{cases}$
\end{definition}

The concept of RQC-model seems to have a ``quasi
intensional'' feature: the meaning $\Qum(\alpha)$ of the
sentence $\alpha$ partially reflects the logical form of $\alpha$.
In fact, the dimension of the Hilbert space where
$\Qum(\alpha)$ ``lives'' depends on the number of
occurrences of atomic sentences in $\alpha$.

\begin{definition}(The atomic complexity of $\alpha$). \nl
$At(\alpha)=
\begin{cases}
1 & \text{if} \: \alpha \: \text{is an atomic sentence}; \\
At(\beta) & \text{if} \: \alpha = \lnot \beta \:\text{or}\:
\alpha = \sqrt{\lnot}\beta;\\ At(\beta)+ At(\gamma)+ 1 & \text{if}
\: \alpha = \beta \land \gamma.
\end{cases}$
\end{definition}

(Recall that: $\Qum(\beta \land \gamma)=
T^{(n,m,1)}(\Qum(\beta)\otimes \Qum(\gamma)\otimes
\Qum(\mathbf f))T^{(n,m,1)}$, if $\Qum(\beta) \in
\otimes^n \C^2$ and $\Qum(\gamma) \in \otimes^m \C^2$).
\begin{lemma}
If $At(\alpha)=n$, then $\Qum(\alpha) \in \frak D(\otimes^{n}\C^2).$
\end{lemma}
\begin{proof}
Straightforward. \end{proof} Given a reversible quantum
computational model $\Qum$, any sentence $\alpha$ has a
natural probability-value, which can be also regarded as its
\emph{extensional meaning} with respect to $\Qum$.
\begin{definition}(The probability-value of $\alpha$ in a model $\Qum$). \nl
$\Prob_{\Qum}(\alpha):=\Prob(\Qum(\alpha)).$
\end{definition}

As we already know, qumixs are naturally preordered by two basic
relations: the strong preorder $\preceq$ and the weak preorder
$\le$. This suggests to introduce two different consequence
relations: the \emph{strong} and the \emph{weak consequence}.

\begin{definition}(Strong and weak consequence in a model $\Qum$). \nl
\begin{enumerate}
\item[1.] A sentence $\beta$ is a \emph{strong consequence in a model}
    $\Qum$ of a  sentence $\alpha$ ($\alpha \Vdash_{\Qum} \beta$)
    iff $\Qum(\alpha) \preceq  \Qum(\beta)$;
\item[2.]A sentence $\beta$ is a \emph{weak consequence in a model}
    $\Qum$ of a  sentence $\alpha$ ($\alpha \Vvdash_{\Qum} \beta$)
    iff $\Qum(\alpha) \le  \Qum(\beta)$.
\end{enumerate}
\end{definition}

The notions of \emph{strong} and \emph{weak truth}, \emph{strong}
and \emph{weak logical consequence}, \emph{strong} and \emph{weak
logical truth} can be now defined in the expected way.

\begin{definition}(Strong and weak truth in a model $\Qum$). \nl
\begin{enumerate}
\item[1.] A sentence $\alpha$ is  \emph{strongly true in a model} $\Qum$
    iff $\mathbf t \Vdash_{ \Qum}\alpha$;
\item[2.]A sentence $\alpha$ is  \emph{weakly true in a model} $\Qum$
    iff $\mathbf t \Vvdash_{ \Qum}\alpha$.
\end{enumerate}
\end{definition}

\begin{definition}(Strong and weak logical consequence). \nl
\begin{enumerate}
\item[1.] A sentence $\beta$ is a \emph{strong logical consequence  of a  sentence $\alpha$} ($\alpha \Vdash \beta$)
    iff for any model $\Qum, \alpha \Vdash_{\Qum}\beta$;
\item[2.] A sentence $\beta$ is a \emph{weak logical consequence} of a sentence $\alpha$ ($\alpha \Vvdash \beta$)
    iff for any model $\Qum, \alpha \Vvdash_{\Qum}\beta.$
\end{enumerate}
\end{definition}

\begin{definition}(Strong and weak logical truth). \nl
\begin{enumerate}
\item[1.] A sentence $\alpha$ is  a \emph{strong logical truth}
    iff for any model $\Qum$, $\alpha$ is strongly true in $\Qum$;
\item[2.]A sentence $\alpha$ is  a \emph{weak logical truth}
    iff for any model $\Qum$, $\alpha$ is weakly true in $\Qum$.
\end{enumerate}
\end{definition}

The strong and the weak logical consequence relations ($\Vdash$
and $\Vvdash$) permit us to characterize semantically two
different forms of \emph{quantum computational logic}. We will
indicate by $\mathbf {^{\sqrt{\lnot}}QCL}$ the logic that is
semantically characterized by the strong logical consequence
relation $\models$. At the same time, the logic that is
characterized by the weak consequence relation will be indicated
by $\mathbf {QCL}$. In other words, we have:
\begin{itemize}
\item \space $\beta$ is a logical consequence of $\alpha$ in the
logic $\mathbf {^{\sqrt{\lnot}}QCL}$ ($\alpha \qclr \beta$) iff
$\beta$ is a strong logical consequence of $\alpha$;
\item \space $\beta$ is a logical consequence of $\alpha$ in the
logic $\mathbf {QCL}$ ($\alpha \models_{\mathbf {QCL}} \beta$) iff
$\beta$ is a weak logical consequence of $\alpha$.
\end{itemize}

Clearly, $\mathbf {^{\sqrt{\lnot}}QCL}$ is a sublogic of $\mathbf
{QCL}$. For:
$$ \alpha \Vdash \beta \:\text{implies}\: \alpha \Vvdash \beta. $$
But not the other way around!

An interesting relation between the two logics $\mathbf
{^{\sqrt{\lnot}}QCL}$ and $\mathbf {QCL}$ is described by the
following theorem:

\begin{theorem}\label{th:duelogiche}\nl
$\alpha
 \models_{\mathbf {QCL}} \beta$ iff $\alpha \land \mathbf t\models_{\mathbf {^{\sqrt{\lnot}}QCL}}
\beta \land \mathbf t$.
\end{theorem}
\begin{proof}
The theorem is a direct consequence of the definition of ${\mathbf
{^{\sqrt{\lnot}}QCL}}$ and ${\mathbf {QCL}}$ and of Theorem
\ref{th:debolepreordine}.
\end{proof}

Let us now turn to the concept of \emph{irreversible quantum
computational model} (shortly, \emph{IQC-model}), where
the ``quasi-intensional'' character of reversible models is lost.
In fact, the interpretation of a sentence in an irreversible
model does not generally reflect the logical form of our sentence:
the meaning of the \emph{whole} does not include the meanings of
the \emph{parts}. In spite of this, we will prove that reversible
and irreversible models turn out to characterize the same logic.
\begin{definition}(IQC-model). \nl
{}\nl
An \emph{IQC-model} of $\mathcal{L}$ is a function
$\Qum^{\C^2}: Form^\mathcal{L} \to \mathfrak D(\C^2)$
(which associates to any sentence $\alpha$ of the language a qumix  of $\C ^2$): \\
$\Qum^{\C^2}(\alpha) :=
\begin{cases}
    P_0  & \text{if} \: \alpha = \mathbf f; \\
    \NOT \, \Qum^{\C^2}(\beta) & \text{if} \: \alpha = \lnot \beta; \\
    \SNOT \, \Qum^{\C^2}(\beta) & \text{if} \: \alpha = \sqrt{\lnot} \beta; \\
    \IAND(\Qum^{\C^2}(\beta),\Qum^{\C^2}(\gamma)) & \text{if} \: \alpha = \beta \land \gamma.
\end{cases}$
\end{definition}

The (strong and weak) notions of \emph{consequence}, \emph{truth},
\emph{logical consequence}, \emph{logical truth} are defined like
in the reversible case, \emph{mutatis mutandis}. We will shortly
speak of \emph{strong irreversible logical consequence} and of
\emph{weak irreversible logical consequence}. The logic that is
determined by the strong irreversible logical consequence will be
indicated by $\mathbf {^{\sqrt{\lnot}}IQCL}$, while $\mathbf
{IQCL}$ will represent the logic determined by the weak
irreversible logical consequence.

We will now prove that $\mathbf {^{\sqrt{\lnot}}QCL}$ and $\mathbf
{^{\sqrt{\lnot}}IQCL}$ are the same logic.

\begin{lemma}\label{le:primoverso}
Let $\Qum$ be a RQC-model and let $\Qum^{\C^2}$ be an IQC-model such that for any atomic
sentence $\mathbf q$: $\:\Qum(\mathbf q)=\Qum^{\C^2}(\mathbf q)$. Then, for any sentence $\alpha\in
Form^{\mathcal L}$:
$$ \Prob(\Qum(\alpha))=\Prob(\Qum^{\C^2}(\alpha)). $$
\end{lemma}
\begin{proof}
The proof is by induction on the \emph{length} (i.e. the number
of connectives) of $\alpha$. \nl (i)\space $\alpha=\mathbf q$. Trivial. \nl
(ii) \space $\alpha=\lnot\beta$.
\begin{align*}
\Prob(\Qum(\alpha)) &= \Prob(\Qum(\lnot\beta)) \\
&=\Prob(\NOT\,\Qum(\beta))\\
&=1-\Prob(\Qum(\beta)) \tag{Theorem \ref{th:guddernot}(ii)}\\
&=1-\Prob(\Qum^{\C^2}(\beta))  \tag{Induction hypothesis}\\
&=\Prob(\NOT\,\Qum^{\C^2}(\beta))\\
&=\Prob(\Qum^{\C^2}(\lnot\beta)).
\end{align*}\nl
(iii)\space $\alpha=\sqrt{\neg}\beta$. The following subcases
are possible: (iiia)\space $\beta=\mathbf q$; (iiib)\space
$\beta=\gamma\land\delta$; (iiic)\space $\beta=\sqrt{\neg}\gamma$;
(iiid)\space $\beta=\neg\gamma$. \nl (iiia)\space $\beta=\mathbf
q$. The proof follows from the assumption $\Qum(\mathbf
q)=\Qum^{\C^2}(\mathbf q)$. \nl (iiib)\space
\begin{align*}
\Prob(\Qum(\alpha))&=\Prob(\Qum(\sqrt{\lnot}\beta) \\
&=\Prob(\SNOT\,\Qum(\gamma\land\delta)) \\
&= \Prob(\SNOT\,\AND(\Qum(\gamma),\Qum(\delta))) \\
&=\um \tag{Theorem \ref{th:guddernot}(v)} \\
\end{align*}
By induction hypothesis and by Lemma \ref{le:iand}(v), we have: $
\Prob(\Qum^{\C^2}(\alpha)) = \Prob(\Qum^{\C^2}(\sqrt{\neg}(\gamma\land\delta)))=
\Prob(\SNOT\,\IAND(\Qum^{\C^2}(\gamma),
\Qum^{\C^2}(\delta)))= \um=\Prob(\Qum(\alpha)). $
\nl (iiic)\space
\begin{align*}
\Prob(\Qum(\alpha))&=\Prob(\Qum(\sqrt{\neg}\sqrt{\neg}\gamma))\\
&=\Prob(\NOT\,\Qum(\gamma))\\
&=1-\Prob(\Qum(\gamma)) \tag{Theorem \ref{th:guddernot}(ii)}\\
&=1-\Prob(\Qum^{\C^2}(\gamma))  \tag{Induction hypothesis}\\
&=\Prob(\NOT\,\Qum^{\C^2}(\gamma))\\
&=\Prob(\SNOT\,\SNOT\,\Qum^{\C^2}(\gamma))\\
&=\Prob(\Qum^{\C^2}(\sqrt{\neg}\sqrt{\neg}\gamma))\\
&=\Prob(\Qum^{\C^2}(\alpha)).
\end{align*}
\nl (iiid)\space The proof follows from induction hypothesis and
Theorem \ref{th:guddernot}(iii). \nl
(iv) \space $\alpha=\beta\land\gamma$.
\begin{align*}
\Prob(\Qum(\alpha))&=\Prob(\Qum(\beta\land\gamma)) \\
&= \Prob(\Qum(\beta))\Prob(\Qum(\gamma)) \tag{Theorem \ref{th:guddernot} (iv)}\\
&= \Prob(\Qum^{\C^2}(\beta))\Prob(\Qum^{\C^2}(\gamma)) \tag{Induction hypothesis}\\
&=\Prob(\IAND(\Qum^{\C^2}(\beta),\Qum^{\C^2}(\gamma))) \tag{Lemma \ref{le:iand} (iv)}\\
&=\Prob(\Qum^{\C^2}(\beta\land\gamma)).
\end{align*}
\end{proof}

\begin{corollary}\label{co:nacht}\nl
\begin{enumerate}
\item[(i)]\space For any RQC-model $\Qum$, there exists an IQC-model
    $\Qum^{\C^2}$ such that for any $\alpha\in Form ^{\mathcal L}$:
$$ \Prob(\Qum(\alpha))=\Prob(\Qum^{\C^2}(\alpha)); $$
\item[(ii)]\space For any IQC-model $\Qum^{\C^2}$ there exists a RQC-model
    $\Qum$ such that for any $\alpha\in Form^{\mathcal L}$:
$$ \Prob(\Qum^{\C^2}(\alpha))=\Prob(\Qum(\alpha)). $$
\end{enumerate}
\end{corollary}

\begin{theorem} \label{th:main}
$\alpha \models_{\mathbf {^{\sqrt{\lnot}}QCL}} \beta$ iff $\alpha
\models_{\mathbf {^{\sqrt{\lnot}}IQCL}} \beta.$
\end{theorem}
\begin{proof}
The theorem is a direct consequence of Corollary \ref{co:nacht}.
\end{proof}

Hence, $\mathbf { ^{\sqrt{\neg}}QCL}$ and $\mathbf {
^{\sqrt{\neg}}IQCL}$ are the same logic. Similarly one can prove
that $\mathbf {QCL}$ and $\mathbf {IQCL}$ are the same logic.

So far we have considered (reversible and irreversible) models,
where the meaning of any sentence is represented by a qumix. A
natural question arises: do density operators have an essential
role in characterizing the logics $\mathbf { ^{\sqrt{\neg}}QCL}$
and $\mathbf {QCL}$? This question has a negative answer. In fact,
one can prove that quregisters are sufficient for our logical aims
in the case of the \emph{minimal} quantum computational language
$\mathcal L$.

Let us first introduce the notion of (reversible) \emph{qubit-model}
(which is the basic concept of the qubit-semantics
described in \cite{CDCGL-02} and \cite{DCGLL-02}).

\begin{definition}(Reversible qubit-model). \nl
A \emph{reversible qubit-model} of $\mathcal{L}$ is a function
$\Qub: Form^\mathcal{L} \to \mathfrak R$
(which associates to any sentence $\alpha$ of the language a quregister): \\
$\Qub(\alpha) :=
\begin{cases}
    \text{a qubit in $\C^2$} & \text{if}\: \alpha\: \text{ is an atomic sentence}; \\
    \ket{0} & \text{if}\: \alpha=\mathbf f;\\
    \QNot(\Qub(\beta)) & \text{if} \: \alpha = \lnot \beta; \\
    \QSNot(\Qub(\beta)) & \text{if} \: \alpha = \sqrt{\lnot} \beta; \\
    \QAnd(\Qub(\beta),\Qub(\gamma)) & \text{if} \: \alpha = \beta \land \gamma.
\end{cases}$
\end{definition}

The notions of (weak and strong) \emph{consequence, truth, logical
consequence, logical truth} are defined like in the case of
reversible qumix models, \emph{mutatis mutandis}. We will write
$\alpha \models_{\mathbf {^{\sqrt{\lnot}}QCL}}^{\Qub}\beta$,
when $\beta$ is a strong logical consequence of
$\alpha$ in the qubit-semantics. Similarly, we will write $\alpha
\models_{\mathbf {QCL}}^{\Qub} \beta$ when $\beta$ is a
weak logical consequence in the same semantics.

Instead of the class $\frak R$ of all quregisters, we could
equivalently refer to the class $\frak {D}_\frak R$ of all
\emph{pure density operators} having the form $P_{\ket{\psi}}$,
where $\ket{\psi}$ is a quregister. One can easily show that
$\frak {D}_\frak R$ is closed under the gates ${\tt
NOT}$,$\sqrt{{\tt NOT}}$,${\tt AND}$. At the same time, $\frak
{D}_\frak R$ is not closed under ${\tt IAND}$, because (as we have
seen) ${\tt IAND}(P_{\ket{\psi}},P_{\ket{\varphi}})$ is,
generally, a proper mixture.

\begin{lemma}\label{le:registripuri}
Consider a reversible qubit-model $\Qub$ and let $\Qum$
be a RQC-model such that for any atomic sentence
$\mathbf q$, $\Qum(\mathbf q)= P_{\Qub(\mathbf
q)}$. Then, for any sentences $\alpha$:
$$ \Qum(\alpha)\approxeq P_{\Qub(\alpha)}. $$
\end{lemma}
\begin{proof}
Easy.
\end{proof}

On this basis we can prove that the qubit-semantics and the
qumix-semantics characterize the same logics.

\begin{theorem} \label{th:bastaqub}\nl
\begin{enumerate}
\item [(1)]$\alpha \models_{\mathbf {^{\sqrt{\lnot}}QCL}} \beta$ iff
$\alpha \models_{\mathbf {^{\sqrt{\lnot}}QCL}}^{\Qub}\beta$;
\item [(2)]$\alpha \models_{\mathbf {QCL}} \beta$ iff
$\alpha \models_{\mathbf {QCL}}^{\Qub}\beta$.
\end{enumerate}
\end{theorem}
\begin{proof}\nl
\begin{enumerate}
\item[(1)]\nl
\begin{enumerate}
\item[(1.1)]\space Suppose that $\alpha \models_{\mathbf
{^{\sqrt{\lnot}}QCL}} \beta$. Then for any RQC-model
$\Qum$:$\Qum(\alpha) \preceq \Qum(\beta)$.Hence, for any $\Qum$ such that
$\Qum(\alpha)$ and $\Qum(\beta)$ are pure density
operators: $\Qum(\alpha) \preceq \Qum(\beta)$.\nl Consequently, by Lemma \ref{le:registripuri},
for any qubit-model $\Qub$:$\Qub(\alpha) \preceq
\Qub(\beta)$. \item[(1.2)]\space Suppose, by
contradiction, that $\alpha \models_{\mathbf
{^{\sqrt{\lnot}}QCL}}^{\Qub}\beta$ and $\alpha
\nvDash_{\mathbf {^{\sqrt{\lnot}}QCL}} \beta$. Then, by Theorem
\ref{th:main} there exists an irreversible model $\mathbf
{Qum^{\C^2}}$ such that $\mathbf {Qum^{\C^2}}(\alpha) \npreceq
\mathbf {Qum^{\C^2}}(\beta)$. By Lemma \ref{le:purissimo}, there
exists a qubit-model $\Qub$ such that for any sentential
letter $\mathbf q$: $\Prob(\Qub(\mathbf q))=
\Prob(\mathbf {Qum^{\C^2}}(\mathbf q))$ and $\Prob(\QSNot(\Qub(\mathbf q)))= \Prob(\SNOT\mathbf
{Qum^{\C^2}}(\mathbf q))$. One can easily prove that for any
$\alpha$, $\Prob(\Qub(\alpha))= \Prob(\mathbf{Qum^{\C^2}}(\alpha))$ and
$\Prob(\QSNot(\Qub(\alpha)))= \Prob(\SNOT\mathbf {Qum^{\C^2}}(\alpha))$
(by induction on the length of $\alpha$). \nl
Consequently, $\alpha \nvDash_{\mathbf {^{\sqrt{\lnot}}QCL}}^{\Qub}\beta$,
contradiction.
\end{enumerate}
\item[(2)] Similarly.
\end{enumerate}
\end{proof}

Needless to observe, Theorem \ref{th:bastaqub} does not imply that
the qumix-semantics is useless. First of all, qubit-models and
qumix-models might characterize different logics for languages
that are richer than $\mathcal L$. At the same time, even in the
case of our minimal language $\mathcal L$, qumixs represent an
important tool in order to describe entanglement-phenomena.

A remarkable property of the logics $\mathbf { ^{\sqrt{\neg}}QCL}$
and $\mathbf {QCL}$ is the following: our logics do not admit any
``genuine'' logical truth. In other words, any sentence $\alpha$,
that does not contain the atomic sentence $\mathbf f$, cannot be a
logical truth. By Theorem \ref{th:bastaqub}, is is sufficient to
prove that no ``genuine'' logical truths exist in the framework of
the qubit-semantics.

Let us first prove the following theorem (\cite{DCGLL-02}):

\begin{theorem}\label{th:threevalue}
Let $\Qub$ be a reversible qubit-model and let $\alpha$ be any sentence.
If $\Prob(\Qub(\alpha)) \in \{ 0,1 \}$, then there is an atomic subformula
$\mathbf q$ of $\alpha$ such that $\Prob(\Qub(\mathbf q)) \in \{ 0,\um,1 \}$.
\end{theorem}
\begin{proof}
 Suppose that $\Prob(\Qub(\alpha)) \in \{ 0,1 \}$. The proof
is by induction on the length of $\alpha$.
\par\noindent
(i)\space $\alpha$ is an atomic sentence. The proof is trivial.
\par\noindent
(ii)\space $\alpha = \lnot \beta$. By Theorem
\ref{th:guddernot}(ii), $\Prob(\Qub(\alpha))=1-\Prob(\Qub(\beta)) \in \{ 0,1 \}$. The
conclusion follows by induction hypothesis.
\par\noindent
(iii)\space $\alpha = \sqrt{\lnot} \beta$. By hypothesis and by
Theorem \ref{th:guddernot}(v), $\beta$ cannot be a conjunction.
Consequently, only the following cases are possible: (iiia)\space
$\beta = \mathbf q$; (iiib)\space $\beta = \lnot \gamma$;
(iiic)\space $\beta = \sqrt{\lnot} \gamma$.
\par\noindent
(iiia)\space $\beta = \mathbf q$. By hypothesis,
$\Prob(\sqrt{\lnot}\beta)\in\{0,1\}$. Hence, $\QSNot
(\Qub(q))=c\ket{x}$, where $\ket{x}\in\{\ket{0},\ket{1}
\}$ and $|c|=1$. We have: \nl$\QNot(\Qub(\mathbf q))=
\QSNot( \QSNot (\Qub(\mathbf
q)))=\QSNot(c\ket{x})$. One can easily show that
$\Prob(\QSNot(c\ket{x})=\frac{1}{2}$. As a consequence,
$\Prob(\Qub(\neg \mathbf q))=\frac{1}{2}=\Prob(\Qub(\mathbf q))$.
\par\noindent
(iiib)\space $\beta = \lnot \gamma$. By Theorem
\ref{th:guddernot}(iii), $\Prob(\Qub(\sqrt{\lnot} \lnot
\gamma))= \Prob(\Qub(\lnot \sqrt{\lnot}
\gamma))=1-\Prob(\Qub(\sqrt{\lnot} \gamma))$. The
conclusion follows by induction hypothesis.
\par\noindent
(iiic)\space $\beta = \sqrt{\lnot} \gamma$.
Then $\Prob(\Qub(\sqrt{\lnot} \sqrt{\lnot} \gamma))= \Prob(\Qub(\lnot \gamma))=1-\Prob(\Qub(\gamma))$.
The conclusion follows by induction hypothesis.
\par\noindent
(iv)\space $\alpha = \beta \land \gamma$. By Theorem
\ref{th:guddernot}(iv), $\Prob(\Qub(\beta\land\gamma))=\Prob(\Qub(\beta)) \Prob(\Qub(\gamma)) \in \{ 0,1 \}$.
The conclusion follows by induction hypothesis.
\end{proof}

As a consequence, we immediately obtain the following Corollary.
\begin{corollary} \label{co:notautologie}
If $\alpha$ does not contain $\mathbf f$, then $\alpha$ is not a logical truth either of
$\mathbf {^{\sqrt{\neg}} QCL}$ or of $\mathbf {QCL}$.
\end{corollary}
\begin{proof}
Suppose, by contradiction, that $\alpha$ is a logical truth either
of $\mathbf {^{\sqrt{\neg}} QCL}$ or of $\mathbf {QCL}$. Then, in
both cases, we obtain that: $\Prob(\alpha)=1$. Let $\mathbf
q_1,\ldots ,\mathbf q_n$ be the atomic sentences occurring in
$\alpha$. Since $\alpha$ does not contain $\mathbf f$, there
exists a qubit-model $\Qub$ such that for any $i$ ($1
\le i \le n$), $\Prob(\Qub(\mathbf q_i))\notin
\{0,\frac{1}{2},1\}$. Then, by Theorem~\ref{th:threevalue},
$\Prob(\Qub(\alpha)) \notin \{0,1\}$, contradiction.
\end{proof}

We will now list some interesting logical consequences and rules
that hold for the logics $\mathbf { ^{\sqrt{\neg}}QCL}$ and
$\mathbf {QCL}$.We will indicate by $\alpha \models \beta$ the
logical consequence relation that refers either to $\mathbf {
^{\sqrt{\neg}}QCL}$ or to $\mathbf {QCL}$. According to the usual
notation we will write:
$$\frac {\alpha_1 \models \beta_1,\ldots,\alpha_n \models \beta_n}{\gamma \models \delta}, $$
to be read as: if $\alpha_1 \models \beta_1,\ldots,\alpha_n
\models \beta_n$, then $\gamma \models \delta$. We will also write
$\alpha \equiv \beta$ as an abbreviation for: $\alpha \models
\beta$ and $\beta \models \alpha$.

Since $\mathbf { ^{\sqrt{\neg}}QCL}$ is a sublogic of $\mathbf
{QCL}$, any logical consequence that holds in $\mathbf {
^{\sqrt{\neg}}QCL}$ will also hold in $\mathbf {QCL}$. At the same
time, some rules that hold in $\mathbf { ^{\sqrt{\neg}}QCL}$ may
be violated in $\mathbf {QCL}$ (and, of course, viceversa).

\begin{theorem}[Logical consequences and rules of both
 $\mathbf {^{\sqrt{\neg}}QCL}$ and $\mathbf {QCL}$] \nl
\begin{enumerate}
\item [(1)]\space  $\alpha \models \alpha;$ \\
    (identity) \\[0.3cm]
\item[(2)]\space $\frac {\alpha \:\models\: \beta,\:\beta\: \models \:\gamma}{\alpha\: \models \: \gamma};$ \\
    (transitivity) \\[0.3cm]
\item[(3)]\space $\alpha \equiv \neg\neg\alpha;$ \\
    (double negation) \\[0.3cm]
\item[(4)]\space $\frac {\alpha\: \models\:\beta}{\neg \beta\:\models\:\neg \alpha};$ \\
    (contraposition for the negation) \\[0.3cm]
\item[(5)]\space $\sqrt{\neg}\sqrt{\neg}\alpha\equiv\neg\alpha;$ \\
    (the double square root of the negation principle) \\[0.3cm]
\item[(6)]\space $\neg\sqrt{\neg}\alpha\equiv \sqrt{\neg}\neg\alpha;$ \\
    (permutation of the negations) \\[0.3cm]
\item[(7)] \space $\sqrt{\neg}\mathbf f \models \sqrt{\neg}\mathbf t;$ \\
    (a ``tentative negation'' of the falsity implies a ``tentative negation'' of the truth) \\[0.3cm]
\item[(8)]\space $\alpha\land\beta\equiv\beta\land\alpha, \quad \alpha\lor\beta\equiv\beta\lor\alpha;$ \\
    (commutativity) \\[0.3cm]
\item[(9)]\space $\alpha\land (\beta\land\gamma)\equiv (\alpha\land\beta)\land\gamma,\quad
    \alpha\lor (\beta\lor\gamma)\equiv(\alpha\lor\beta)\lor\gamma;$ \\
    (associativity) \\[0.3cm]
\item[(10)]\space $\neg(\alpha\land\beta)\equiv\neg\alpha\lor\neg\beta,
    \quad  \neg(\alpha\lor\beta)\equiv\neg\alpha\land\neg\beta;$ \\
    (de Morgan) \\[0.3cm]
\item[(11)]\space $\alpha\land(\beta\lor\gamma)\models(\alpha\land\beta)\lor(\alpha\land\gamma), \quad
    (\alpha\lor\beta)\land(\alpha\lor\gamma)\models\alpha\lor(\beta\land\gamma);$ \\
    (distributivity 1) \\[0.3cm]
\item[(12)]\space $\mathbf f \land \mathbf f \equiv \mathbf f,\quad \mathbf t\land \mathbf t\equiv \mathbf t;$ \\
    (idempotence for the truth and the falsity) \\[0.3cm]
\item[(13)]\space $\mathbf f \land \mathbf t \equiv \mathbf f,\quad\mathbf f \lor \mathbf t\equiv \mathbf t;$ \\[0.3cm]
\item[(14)]\space $\frac{\alpha\quad \equiv\quad\beta}{\neg\alpha \quad\equiv\quad \neg\beta};$ \\
    (logical equivalence is a congruence for the negation) \\[0.3cm]
\item[(15)]$\frac{\alpha  \quad \equiv \quad \gamma,\quad\beta\quad \equiv \quad \delta}{\alpha\land \beta\quad
    \equiv \quad \gamma \land \delta};$ \\
    (logical equivalence is a congruence for the conjunction) \\[0.3cm]
\item[(16)]\space $\sqrt{\neg}(\alpha \land \beta) \models \sqrt{\neg}\mathbf t;$ \\[0.3cm]
\item[(17)]\space $\frac{\sqrt{\neg} \alpha\quad \models\quad \sqrt{\neg}\mathbf t}{\alpha \land \beta
    \quad\models\quad \alpha}, \quad \frac{\sqrt{\neg} \beta
    \quad\models\quad \sqrt{\neg}\mathbf t}{\alpha \land \beta\quad \models \quad\beta};$ \\[0.3cm]
\item[(18)]\space $\frac{\alpha \quad \models \quad \sqrt{\neg} \mathbf t} {\mathbf f \models \alpha}.$ \\
    (Weak Duns Scotus)
\end{enumerate}
\end{theorem}
\begin{proof}
Easy.
\end{proof}

Let us now consider examples of logical consequences and rules
that hold in $\mathbf {QCL}$ and are violated in $\mathbf
{^{\sqrt{\neg}}QCL}$.
\begin{theorem}[Logical consequences and rules of $\mathbf {QCL}$
that fail in
$\mathbf {^{\sqrt{\neg}}QCL}$]\nl
\begin{enumerate}
\item[(1)]\space $\alpha \land \beta \models_{\mathbf{QCL}} \alpha,
    \quad \alpha \land \beta \models_{\mathbf{QCL}}\beta;$ \\[0.3cm]
\item[(2)] \space $\alpha  \models_{\mathbf{QCL}}\alpha \lor \beta,
    \quad \beta \models_{\mathbf{QCL}}\alpha \lor \beta;$ \\[0.3cm]
\item[(3)]\space $\alpha\land\alpha\models_{\mathbf{QCL}}\alpha,
    \quad \alpha \models _{\mathbf{QCL}} \alpha \lor \alpha;$ \\
    (semiidempotence 1) \\[0.3cm]
\item[(4)] $\mathbf f \models _{\mathbf{QCL}} \alpha.$ \\
(Duns Scotus)
\end{enumerate}
\end{theorem}
\begin{proof}
Easy.
\end{proof}

\begin{theorem}[A rule that holds in $\mathbf {^{\sqrt{\neg}}
QCL}$ and fails in $\mathbf {QCL}$] \nl
$$\frac{\alpha \equiv \beta}{\sqrt{\neg}\alpha \equiv \sqrt{\neg}\beta}. $$
\end{theorem}
\begin{proof}
Easy.
\end{proof}

In other words, logical equivalence is a congruence for the square
root of the negation.

\begin{theorem}[Logical consequences that fail both in
$\mathbf {QCL}$ and $\mathbf {^{\sqrt{\neg}} QCL}$]\nl
\begin{enumerate}
\item[(1)]\space  $\alpha\not\models\alpha\land\alpha;$ \\
    (semiidempotence 2)
\item[(2)]\space $\mathbf t \not\models\alpha\lor\neg\alpha;$ \\
    (excluded middle)
\item[(3)]\space $\mathbf t \not\models\neg(\alpha\land\neg\alpha);$ \\
(non contradiction)
\item[(4)]\space $(\alpha\land \beta)\lor(\alpha\land\gamma) \not\models \alpha\land(\beta\lor\gamma),
    \quad \alpha\lor(\beta\land\gamma) \not\models (\alpha\lor \beta)\land(\alpha\lor\gamma).$ \\
    (distributivity 2)
\end{enumerate}
\end{theorem}
\begin{proof}
Easy.
\end{proof}

Apparently, the logics $\mathbf {QCL}$ and $\mathbf {^{\sqrt
{\neg}}QCL}$ turn out to be non standard forms of quantum logic.
Conjunction and disjunction do not correspond to lattice
operations, because they are not generally idempotent. Unlike
Birkhoff and von Neumann's quantum logic, the weak distributivity
principle ($(\alpha\land\beta)\lor(\alpha\land\gamma)\models
\alpha\land(\beta\lor\gamma)$) breaks down.
At the same time, the strong distributivity
($\alpha\land(\beta\lor\gamma)\models(\alpha\land\beta)\lor(\alpha\land\gamma)$),
that is violated in orthodox quantum logic, is here valid.
Both the excluded middle and the non contradiction principles are
violated.As a consequence, one can say that the logics arising
from quantum computation represent, in a sense, new examples of
\emph{fuzzy logics}.

The axiomatizability of $\mathbf {QCL}$ and $\mathbf {^{\sqrt
{\neg}}QCL}$ is an open problem.

\section{Quantum trees}

An interesting feature of the quantum computational semantics is
the following: the \emph{meaning} and the probability-value of any
molecular sentence $\alpha$ can be naturally described (and
calculated) by means of a convenient \emph{quantum tree}, that
illustrates a kind of reversible transformation of the atomic
subformulas of $\alpha$. By theorem \ref{th:bastaqub}, we know
that we can refer to the qubit-semantics (instead of the
qumix-semantics), without any loss of generality. For the sake of
technical simplicity, we will first slightly modify our language.
The new language $\mathcal L^{\bigwedge}$ contains, besides the
atomic sentence $\mathbf f$ and the two negations ($\lnot$ and
$\sqrt{\lnot}$), a ternary conjunction $\bigwedge$ (whose semantic
behaviour is ``close'' to the Petri-Toffoli gate). For any
sentences $\alpha$ and $\beta$, the expression
$\bigwedge(\alpha,\beta,f)$ is a sentence of $\mathcal
L^{\bigwedge}$. In this framework, the usual conjunction $\alpha
\land \beta$ is dealt with as meta\-linguistic abbreviation for
the ternary conjunction $\bigwedge(\alpha, \beta,\mathbf f)$. The
occurrence of $\mathbf f$ as the third element in the formula
$\bigwedge(\alpha, \beta,\mathbf f)$ is called a
\emph{non-genuine occurrence} of $\mathbf f$. The semantic definition
of \emph{qubit-model} of the language $\mathcal L^{\bigwedge}$ is
then modified in the expected way. Besides the old conditions
concerning the interpretation of $\mathbf f$ and of the two
negations ($\neg$, $\sqrt{\neg}$), we require that for any
$\Qub$:
$$ \Qub(\bigwedge(\alpha,\beta,\mathbf f))
    =T(\Qub(\alpha),\Qub(\beta),\Qub(\mathbf f)). $$

Needless to stress, the logics $\mathbf {QCL}$ and $\mathbf
{^{\sqrt {\neg}}QCL}$ can be equivalently formalized either in the
language $\mathcal L$ or in $\mathcal L^{\bigwedge}$. In case
where the language is $\mathcal L^{\bigwedge}$, Corollary
\ref{co:notautologie} shall be formulated as follows: \emph{if
$\alpha$ does not contain any genuine occurrence of $\mathbf f$,
then $\alpha$ is not a logical truth either of $\mathbf
{^{\sqrt{\neg}} QCL}$ or of $\mathbf {QCL}$}.

Before dealing with quantum trees, we will first introduce the
notion of \emph{syntactical tree} of a sentence $\alpha$
(abbreviated as $STree^{\alpha}$). Consider all subformulas of
$\alpha$.

Any subformula may be:
\begin{itemize}
\item an \emph{atomic} sentence $\mathbf q$ (possibly $\mathbf f$);
\item a \emph{negated}  sentence $\lnot \beta$;
\item a \emph{square-root negated} sentence $\sqrt{\lnot}\beta$;
\item a \emph{conjunction} $\bigwedge(\beta,\gamma,\mathbf f)$.
\end{itemize}

\begin{figure}[t]
\begin{center}
\includegraphics[width=8.5cm]{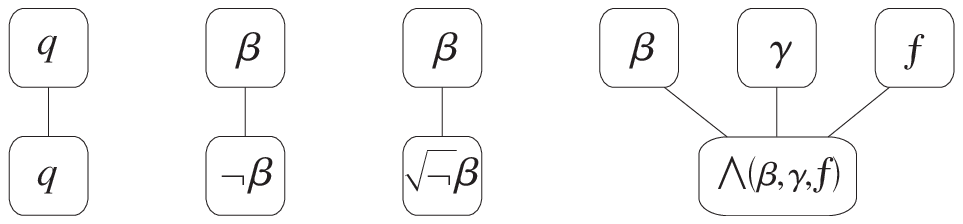}
\end{center}
\caption\emph{Branching rules for the construction of syntactical
trees.} \label{fig:branching-rules}
\end{figure}

The intuitive idea of \emph{syntactical tree} can be illustrated
as follows. Every occurrence of a subformula of $\alpha$ gives
rise to a \emph{node} of $STree^{\alpha}$. The tree consists of a
finite number of \emph{levels} and each level is represented by a
sequence of subformulas of $\alpha$:
\vspace{-0.5cm}
$$ Level_k(\alpha) $$
\vspace{-0.5cm}
$$ \vdots $$
\vspace{-0.5cm}
$$ Level_1(\alpha). $$
The \emph{root-level} (denoted by
$Level_1(\alpha)$) consists of $\alpha$. From each node of the
tree at most 3 edges may branch according to the \emph{branching-rule}
(Figure \ref{fig:branching-rules}).

The second level ($Level_2(\alpha)$) is the sequence of
subformulas of $\alpha$ that is obtained by applying the
branching-rule to $\alpha$. The third level ($Level_3(\alpha)$) is
obtained by applying the branching-rule to each element (node) of
$Level_2(\alpha)$, and so on. Finally, one obtains a level
represented by the sequence of all atomic occurrences of $\alpha$.
This represents the \emph{last level} of $STree^{\alpha}$. The
\emph{height} of $Stree^{\alpha}$ (denoted by $Height(\alpha)$)
is then defined as the number of levels of $STree^{\alpha}$.

A more formal definition of \emph{syntactical tree} can be given
by using some standard graph-theoretical notions.

\begin{example}
The syntactical tree of
$\alpha = \lnot \mathbf q\land (\mathbf r \land \sqrt{\lnot} \mathbf q)$ is the following:
\begin{figure}[h]
\begin{center}
    \includegraphics[width=9.5cm]{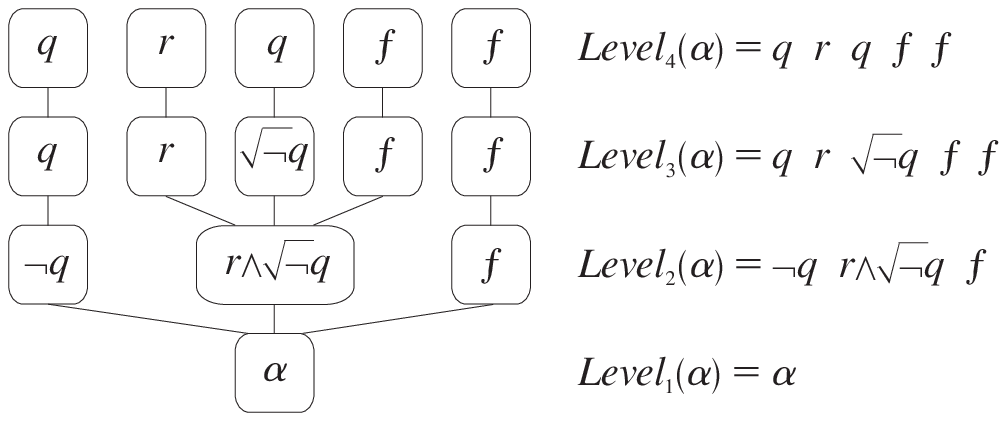}
\end{center}
\end{figure}
Clearly the height of $Stree^{\alpha}$ is 4.
\end{example}

For any choice of a qubit-model $\Qub$, the syntactical
tree of $\alpha$ determines a corresponding sequence of
quregisters. Consider a sentence $\alpha$ with $n$ atomic
occurrences ($\mathbf q_1,\ldots,\mathbf q_n$). Then $\Qub(\alpha) \in \otimes^n \C^2$. We can associate a quregister
$\ket{\psi_i}$ to each $Level_i(\alpha)$ of $Stree^{\alpha}$ in
the following way. Suppose that:
$$ Level_i(\alpha)=(\beta_1,\ldots, \beta_r). $$
Then:
$$ \ket{\psi_i}=\Qub(\beta_1)\otimes \ldots \otimes \Qub(\beta_r). $$
Hence:
$$
\begin{cases}
    \ket{\psi_1}=\Qub(\alpha) \\
    \vdots \\
    \ket{\psi_{Height(\alpha)}}=\Qub(\mathbf q_1) \otimes \ldots \otimes \Qub(\mathbf q_n)
\end{cases}
$$
where all $\ket{\psi_i}$ belong to the same space $\otimes^n
\C^2$.

From an intuitive point of view, $\ket{\psi_{Height(\alpha)}}$ can
be regarded as a kind of \emph{epistemic state}, corresponding
to the input of a computation, while $\ket{\psi_1}$ represents the
output.

We obtain the following correspondence:

\begin{align*}
\mbox{$Level_{Height(\alpha)}(\alpha)$} & \mbox{$\leftrightsquigarrow \ket{\psi_{Height(\alpha)}}$: the input} \\
\mbox{$\ldots$} & \mbox{$\leftrightsquigarrow \ldots$} \\
\mbox{$Level_1(\alpha)$} & \mbox{$\leftrightsquigarrow \ket{\psi_1}$: the output}
\end{align*}

The notion of \emph{quantum tree} of a sentence $\alpha$
 ($QTree^{\alpha})$ can be now defined as a particular sequence of
unitary operators that is uniquely determined by the syntactical
tree of $\alpha$. As we already know, each $Level_i(\alpha)$ of
$STree^{\alpha}$ is a sequence of subformulas of $\alpha$. Let
$Level_i^j(\alpha)$ represent the $j$-th element of
$Level_i(\alpha)$. Each node $Level_i^j(\alpha)$ (where $1 \le i
< Height(\alpha)$) can be naturally associated to a unitary
operator $Op^j_i$, according to the following \emph{operator-rule}:
\[
Op^j_i :=
\begin{cases}
I^{(1)} & \text{if $Level_i^j(\alpha)$ is an atomic sentence}; \\
\QNot^{(r)} & \text{if $Level_i^j(\alpha)=\lnot \beta$ and $\Qub(\beta) \in \otimes^r \C^2$}; \\
\sqrt{\QNot}^{(r)} & \text{if $Level_i^j(\alpha)=\sqrt{\lnot} \beta$ and $\Qub(\beta) \in \otimes^r \C^2$}; \\
T^{(r,s,1)} & \text{if $Level_i^j(\alpha)=
    \bigwedge(\beta,\gamma,\mathbf f)$, $\Qub(\beta) \in \otimes^r \C^2$ and
    $\Qub(\gamma) \in \otimes^s \C^2$}.
\end{cases}
\]

On this basis, one can associate an operator $U_i$ to each
$Level_i(\alpha)$ (such that $1 \le i < Height(\alpha)$):
$$ U_i := \bigotimes_{j=1}^{|Level_i(\alpha)|} Op_i^j, $$
where $|Level_i(\alpha)|$ is the length of the sequence $Level_i(\alpha)$.

Being the tensor product of unitary operators, every $U_i$ turns
out to be a unitary operator. One can easily show that all $U_i$
are defined in the same space $\otimes^n \C^2$, where $n$ is the
atomic complexity of $\alpha$.

The notion of \emph{quantum tree} of a sentence can be now
defined as follows.
\begin{definition}(The quantum tree of $\alpha$). \nl
The \emph{quantum tree} of $\alpha$ (denoted by $QTree^{\alpha}$)
is the operator-sequence
$$ (U_1,\ldots,U_{Height(\alpha)-1}) $$
that is uniquely determined by the syntactical tree of $\alpha$.
\end{definition}

As an example, consider the following sentence: $\alpha=\mathbf q
\land \lnot \mathbf q= \bigwedge(\mathbf q,\lnot \mathbf q,\mathbf
f)$. The syntactical tree of $\alpha$ is the following:
\begin{align*}
\mbox {$Level_1(\alpha)= \bigwedge(\mathbf q,\lnot \mathbf q,\mathbf f);$}\\
\mbox {$Level_2(\alpha)= (\mathbf q,\lnot \mathbf q,\mathbf f);$}\\
\mbox {$Level_3(\alpha)= (\mathbf q,\mathbf q,\mathbf f).$}
\end{align*}

In order to construct the quantum tree of $\alpha$, let us first
determine the operators $Op^j_i$ corresponding to each node of
$Stree^{\alpha}$. We will obtain:

\begin{itemize}
\item $Op^1_1=T^{(1,1,1)}$, because $\bigwedge(\mathbf q,\lnot
\mathbf q,\mathbf f)$ is connected with $(\mathbf q,\lnot \mathbf
q,\mathbf f)$ (at $Level_2(\alpha)$);
\item $Op_2^1=I^{(1)}$, because $\mathbf q$ is connected with $\mathbf q$ (at $Level_3(\alpha)$);
\item $Op_2^2=\QNot^{(1)}$, because $\lnot \mathbf q$ is connected with $\mathbf q$ (at $Level_3(\alpha)$);
\item $Op_2^3=I^{(1)}$, because $\mathbf f$ is connected with $\mathbf f$ (at $Level_3(\alpha)$).
\end{itemize}

The quantum tree of $\alpha$ is represented by the
operator-sequence $(U_1,U_2)$, where:
\begin{align*}
&U_1= Op_1^1=T^{(1,1,1)}; \\
&U_2= Op_2^1 \otimes Op_2^2 \otimes Op_2^3=I^{(1)} \otimes \QNot^{(1)} \otimes I^{(1)}.
\end{align*}

Apparently, $QTree^{\alpha}$ is independent of the choice of $\Qub$.

\begin{theorem}
Let $\alpha$ be a sentence whose quantum tree
is the operator-sequence $(U_1,\ldots,U_{Height(\alpha)-1})$.
Given a quantum computational model $\Qub$, consider the
quregister-sequence  $(\ket{\psi_1},\ldots,\ket{\psi_{Height(\alpha)}})$
that is determined by $\Qub$ and by the syntactical tree of $\alpha$.
Then, $U_i (\ket{\psi_{i+1}})= \ket{\psi_i}$ (for any $i$ such
that $1 \leq i < Height(\alpha)$).
\end{theorem}
\begin{proof}
Straightforward.
\end{proof}

The quantum tree of $\alpha$ can be naturally regarded as a
\emph{quantum circuit} that computes the output
$\Qub(\alpha)$, given the input $\Qub(\mathbf
q_1),\ldots, \Qub(\mathbf q_n)$ (where $\mathbf
q_1,\ldots,\mathbf q_n$ are the atomic occurrences of $\alpha$).
In this framework, each $U_i$ is the unitary operator that
describes the computation performed by the $i$-th layer of the
circuit.

\end{document}